\documentclass[12pt,letterpaper]{article}
\usepackage{amsmath,amssymb,array,calc,rotating,epsfig,psfrag,amscd, cite}

\makeatletter
\renewcommand\section{\@startsection {section}{1}{\z@}%
                                   {-3.5ex \@plus -1ex \@minus -.2ex}
                                   {2.3ex \@plus.2ex}%
                                   {\normalfont\large\bfseries}}
\renewcommand\subsection{\@startsection{subsection}{2}{\z@}%
                                     {-3.25ex\@plus -1ex \@minus -.2ex}%
                                     {1.5ex \@plus .2ex}%
                                     {\normalfont\bfseries}}
\makeatother

\let\non\nonumber

\let\a=\alpha\let\b=\beta
\let\l=\lambda
\let\r=\rho
\let\s=\sigma

\newcommand{\bea}{\begin{eqnarray}}
\newcommand{\eea}{\end{eqnarray}}
\newcommand{\be}{\begin{equation}}
\newcommand{\ee}{\end{equation}}

\newcommand{\m}{\mu}
\newcommand{\n}{\nu}
\newcommand{\p}{\partial}

\newcommand{\C}[1]{$(\ref{#1})$}

\def\IZ{\relax\ifmmode\mathchoice
{\hbox{\cmss Z\kern-.4em Z}}{\hbox{\cmss Z\kern-.4em Z}}
{\lower.9pt\hbox{\cmsss Z\kern-.4em Z}} {\lower1.2pt\hbox{\cmsss
Z\kern-.4em Z}}\else{\cmss Z\kern-.4em Z}\fi}
\def\IR{\relax{\rm I\kern-.18em R}}

\def\one{{\hbox{ 1\kern-.8mm l}}}

\newlength{\bredde}
\def\slash#1{\settowidth{\bredde}{$#1$}\ifmmode\,\raisebox{.15ex}{/}
\hspace*{-\bredde} #1\else$\,\raisebox{.15ex}{/}\hspace*{-\bredde}
#1$\fi}

\newsavebox{\zzzbar}
\sbox{\zzzbar}
  {\setlength{\unitlength}{0.9em}
  \begin{picture}(0.6,0.7)
  \thinlines
  \put(0,0){\line(1,0){0.6}}
  \put(0,0.75){\line(1,0){0.575}}
  \multiput(0,0)(0.0125,0.025){30}{\rule{0.3pt}{0.3pt}}
  \multiput(0.2,0)(0.0125,0.025){30}{\rule{0.3pt}{0.3pt}}
  \put(0,0.75){\line(0,-1){0.15}}
  \put(0.015,0.75){\line(0,-1){0.1}}
  \put(0.03,0.75){\line(0,-1){0.075}}
  \put(0.045,0.75){\line(0,-1){0.05}}
  \put(0.05,0.75){\line(0,-1){0.025}}
  \put(0.6,0){\line(0,1){0.15}}
  \put(0.585,0){\line(0,1){0.1}}
  \put(0.57,0){\line(0,1){0.075}}
  \put(0.555,0){\line(0,1){0.05}}
  \put(0.55,0){\line(0,1){0.025}}
  \end{picture}}

\newcommand{\ena}{\end{eqnarray}}
\newcommand{\beqa}{\begin{eqnarray}}
\newcommand{\eeqa}{\end{eqnarray}}

\renewcommand{\b}{\beta}

\def\a{\alpha}
\def\b{\beta}

\def\l{\lambda}
\def\m{\mu}
\def\n{\nu}

\def\r{\rho}
\def\s{\sigma}

\begin{document}
\begin{titlepage}

\begin{center}



\vskip 2 cm
{\Large \bf Non--analytic terms from nested divergences in maximal supergravity}\\
\vskip 1.25 cm { Anirban Basu\footnote{email address:
    anirbanbasu@hri.res.in} } \\
{\vskip 0.5cm Harish--Chandra Research Institute, Chhatnag Road, Jhusi,\\
Allahabad 211019, India\\}

\end{center}

\vskip 2 cm

\begin{abstract}
\baselineskip=18pt

The $D^4\mathcal{R}^4$ and $D^6\mathcal{R}^4$ coefficient functions in the effective action of type II string theory compactified on $T^d$ contain terms of the form $\mathcal{E}_1 {\rm ln}g_d$ and $\mathcal{E}_2 ({\rm ln}g_d)^2$ in specific dimensions, where $g_d$ is the T--duality invariant string coupling, and $\mathcal{E}_1$ and $\mathcal{E}_2$ are U--duality invariant coefficient functions. We derive these non--analytic terms from nested ultraviolet divergences in two and three loop maximal supergravity. For the $D^4\mathcal{R}^4$ coupling, the contribution involves $\mathcal{E}_{\mathcal{R}^4} {\rm ln} g_d$, while for the $D^6\mathcal{R}^4$ coupling, it involves $\mathcal{E}_{\mathcal{R}^4} {\rm ln} g_d$, $\mathcal{E}_{D^2\mathcal{R}^4} ({\rm ln} g_d)^2$ and $\mathcal{E}_{D^4\mathcal{R}^4} {\rm ln} g_d$; where $\mathcal{E}_{\mathcal{R}^4}$, $\mathcal{E}_{D^2\mathcal{R}^4}$ and $\mathcal{E}_{D^4\mathcal{R}^4}$ are the $\mathcal{R}^4$, $D^2\mathcal{R}^4$ and $D^4\mathcal{R}^4$ coefficient functions respectively. The contribution from $\mathcal{E}_{D^2\mathcal{R}^4}$, the coefficient function of an amplitude that vanishes onshell, arises from a two loop nested subdivergence of the three loop amplitude.

\end{abstract}

\end{titlepage}


\section{Introduction}

Among the various terms in the effective action of maximally supersymmetric string theories, the BPS protected ones are amenable to a detailed analysis. In particular, among such terms the $\mathcal{R}^4$, $D^4\mathcal{R}^4$ and $D^6\mathcal{R}^4$ terms in the effective action of toroidally compactified type II string theory are BPS protected. These interactions which involve the four graviton amplitude at the linearized level, have been analyzed using various techniques including string perturbation theory, spacetime supersymmetry, U--duality and multiloop supergravity~\cite{Green:1997tv,Green:1997as,Kiritsis:1997em,Green:1998by,Green:1999pu,D'Hoker:2005jc,D'Hoker:2005ht,Berkovits:2005ng,Green:2005ba,Berkovits:2006vc,Basu:2007ru,Basu:2007ck,Green:2008bf,Basu:2008cf,Green:2010kv,Green:2010wi,Basu:2011he,D'Hoker:2013eea,Gomez:2013sla,D'Hoker:2014gfa,Basu:2014hsa,Bossard:2014aea,Pioline:2015yea,Bossard:2015uga,Bossard:2015oxa,Basu:2015dqa,Pioline:2015nfa,Bossard:2015foa}. In the Einstein frame, the moduli dependent coefficient functions of these interactions are U--duality invariant. Along with terms that are analytic in the T--duality invariant string coupling when expanded at weak coupling, these coefficient functions also include terms that are non--analytic in the string coupling in certain dimensions. For toroidal compactification on $\mathbb{R}^{9-d,1}\times T^d$, these contributions which are logarithmic in the string coupling for these BPS interactions, are argued to be given by~\cite{Green:2010sp,D'Hoker:2014gfa,Pioline:2015yea}    
\bea \label{value}\mathcal{E}_{\mathcal{R}^4}^{non-an} &=& \frac{4\pi}{3} {\rm ln} g_2 \delta_{D,8}, \non \\ \mathcal{E}_{D^4 \mathcal{R}^4}^{non-an} &=& \frac{16\pi^2}{15}{\rm ln} g_3 \delta_{D,7} +\mathcal{E}_{\mathcal{R}^4} {\rm ln} g_4 \delta_{D,6}, \non \\ \mathcal{E}_{D^6 \mathcal{R}^4}^{non-an} &=& 5\zeta(3){\rm ln} g_4 \delta_{D,6} + \Big[ -\frac{4\pi^2}{27} ({\rm ln} g_2)^2 +\frac{2\pi}{9} \Big(\frac{\pi}{2} + \mathcal{E}_{\mathcal{R}^4}\Big){\rm ln} g_2\Big]\delta_{D,8} \non \\ &&+ \frac{20}{9} \mathcal{E}_{\mathcal{R}^4} {\rm ln} g_5 \delta_{D,5} +\frac{5}{\pi} \mathcal{E}_{D^4\mathcal{R}^4} {\rm ln} g_6 \delta_{D,4}\eea
based on constraints due to U--duality and terms arising from the boundary of moduli space in string amplitudes.  
Here $g_d$ is the T--duality invariant dilaton defined as $g_d^{-2} = e^{-2\phi_d} = e^{-2\phi} V_d$, where $V_d$ is the volume of $T^d$ in the string frame metric. Also 
\be D=10-d \ee
denotes the number of non--compact $\mathbb{R}^{9-d,1}$ dimensions. We have denoted the U--duality invariant coefficient functions of the  $\mathcal{R}^4$, $D^4\mathcal{R}^4$ and $D^6\mathcal{R}^4$ terms as $\mathcal{E}_{\mathcal{R}^4}$, $\mathcal{E}_{D^4\mathcal{R}^4}$ and $\mathcal{E}_{D^6\mathcal{R}^4}$ respectively.  

Our aim is to understand the origin of some of these terms from maximal supergravity. Very schematically, maximal supergravity has ultraviolet divergences involving ${\rm \ln}(-S/\Lambda^2)$ for these BPS interactions, where $\Lambda$ is the UV cutoff. Here $S$ is a generic Mandelstam variable and we have been schematic about the precise spacetime structure of the logarithmic contribution. In the corresponding string amplitude which is ultraviolet finite, these show up as infrared divergences in the string frame involving ${\rm \ln}(-\mu \alpha' S)$ from the boundary of moduli space, where $\mu$ is a constant that can be calculated directly. On converting to the Einstein frame, this yields a local contribution of the form ${\rm ln} g_d$ coming from the dilaton dependence of $S$ from the inverse metric on converting from the string frame to the Einstein frame. Naturally the scale of the logarithm is ambiguous, however the overall coefficient of the term is unambiguously defined. We choose a convention such that ${\rm \ln}(-\mu \alpha' S)$ contributes ${\rm ln} g_d$ to the local interaction upto an overall factor.       

The various ultraviolet divergences in maximal supergravity which are power behaved in the ultraviolet cutoff $\Lambda$ have been calculated using momentum cutoff and then regularized using symmetries of string theory to define the amplitudes in quantum supergravity (hence the answers are moduli dependent unlike ordinary supergravity, as it goes beyond dimensional reduction). However the logarithmic divergences are easier to calculate using dimensional regularization and isolating the contributions from the $\epsilon$ poles, which is going to be the technique we shall use.

The logarithmic divergences in supergravity for the $\mathcal{R}^4$, $D^4\mathcal{R}^4$ and $D^6\mathcal{R}^4$ terms in 8, 7 and 6 dimensions in \C{value}  appear at one, two and three loops respectively as primitive divergences. This structure follows simply from power counting, and the moduli independent coefficients of these logarithms can indeed be calculated from supergravity loops directly. The divergence for the $D^6\mathcal{R}^4$ term in 8 dimensions arises from $1/\epsilon$ and $1/\epsilon^2$ pole contributions at two loops, on including the one loop $\mathcal{R}^4$ counterterm as well. The moduli dependent part involving $\mathcal{E}_{\mathcal{R}^4}$ arises from a one loop subdivergence\footnote{This yields a $1/\epsilon^2$ term in \C{value} as well, which follows from the first equation in \C{value}. Thus the total coefficient of this double pole term is $+4\pi^2/27$ in ordinary supergravity.}. 

While there are several contributions in \C{value} which simply involve a numerical factor multiplying the logarithms as mentioned above, there are others where the logarithms are multiplied by U--duality invariant coefficient functions, which form the primary focus of our analysis. These include the contributions to the $D^4\mathcal{R}^4$ term in 6 dimensions, and to the $D^6\mathcal{R}^4$ term in 8, 5 and 4 dimensions. These divergences arise from nested one and two loop ultraviolet divergences in two and three loop maximal supergravity which we evaluate using dimensional regularization. This automatically fixes the structure of the moduli dependent U--duality invariant coefficient functions of these logarithms. In order to fix the precise normalization factors for these contributions, we have to know the exact relation between the $\epsilon$ pole in dimensional regularization and the coefficient of the logarithm. This relation is obtained at the required loop orders by directly evaluating logarithmically divergent appropriate field theory amplitudes. 

Now the expressions involving moduli dependent couplings of the logarithms in \C{value} have been argued on the basis of U--duality and the perturbative structure has also been obtained using the low momentum expansion of the four graviton amplitude at various genera. Our results precisely reproduce the terms in \C{value} except for an additional contribution to the non--analytic part of the $D^6\mathcal{R}^4$ amplitude given by
\be \label{add}\mathcal{E}^{non-an}_{D^6\mathcal{R}^4} = - \frac{25\delta_{D,4}}{4\pi^2}\mathcal{E}_{D^2\mathcal{R}^4} ({\rm ln} g_6)^2 \delta_{D,4},\ee
where $\mathcal{E}_{D^2\mathcal{R}^4}$ is the coefficient function of the $D^2\mathcal{R}^4$ amplitude. We argue based on our calculations that even though this amplitude vanishes on--shell, its coefficient function contributes to the $D^6\mathcal{R}^4$ amplitude. This provides a direct derivation of these non--analytic terms in the string coupling from supergravity amplitudes.            

We begin with a discussion of the $\mathcal{R}^4$, $D^4\mathcal{R}^4$ and $D^6\mathcal{R}^4$ interactions that are obtained from one, two and three loop four graviton amplitudes in maximal supergravity, along with the one loop $D^2\mathcal{R}^4$ interaction. Then we perform the detailed analysis of the various logarithmically divergent contributions that arise from nested divergences in two and three loop supergravity. These divergent contributions given by simple poles in $\epsilon$ in dimensional regularization, have coefficient functions that involve the $\mathcal{R}^4$ and $D^4\mathcal{R}^4$ amplitudes. On the other hand, the contribution yielding a double pole involves the $D^2\mathcal{R}^4$ amplitude. These lead to a dependence of the $D^4\mathcal{R}^4$ and $D^6\mathcal{R}^4$ interactions on ${\rm ln}g_d$ with moduli dependent coefficients in specific dimensions given by \C{value} and \C{add}. Though we have focused on specific BPS amplitudes which are the simplest to analyze, the primary logic generalizes to other BPS amplitudes lying in the supermultiplets of the ones we have considered, and to their non--BPS counterparts as well.

\section{The structure of four graviton loop amplitudes in maximal supergravity}

The $\mathcal{R}^4$, $D^4\mathcal{R}^4$ and $D^6\mathcal{R}^4$ terms receive contributions upto one, two and three loops in supergravity respectively. Their contribution to the coefficient functions of these interactions in string theory are calculated by considering these terms in $\mathcal{N}=1$, $d=11$ supergravity compactified on $\mathbb{R}^{9-d,1}\times T^{d+1}$ and using the duality between M theory on $T^{d+1}$ and type II string theory on $T^d$. This yields exact U--duality invariant answers only for $d=0,1$, as compactifications for larger $d$ also include non--perturbative states which are not included in the supergravity approximation. Even then supergravity yields useful insight into the exact answer which will be useful for our purposes. This is because the exact answer $\mathcal{E}$ for any coefficient function is given by
\be \mathcal{E} = \mathcal{E}^{sugra} +\ldots,\ee   
where $\mathcal{E}^{sugra}$ is the answer from supergravity. Thus in our analysis we shall obtain expressions involving $\mathcal{E}^{sugra}$ which can then be completed to $\mathcal{E}$ in a U--duality invariant way, leading to the complete expression for the logarithmic divergences. Since we shall need explicit expressions for the $\mathcal{R}^4$, $D^4\mathcal{R}^4$ and $D^6\mathcal{R}^4$ amplitudes to calculate the various logarithmically divergent contributions in \C{value} and \C{add}, we briefly discuss them below~\cite{Green:1982sw,Green:1997as,Russo:1997mk,Green:1997ud,Bern:1998ug,Green:1999pu,Green:2005ba,Bern:2007hh,Green:2008bf,Bern:2008pv,Basu:2014hsa}.  

In the various calculations that we perform, we often express the propagators in the Schwinger representation. Thus in a diagram involving $n+1$ propagators, we introduce $n+1$ Schwinger parameters $\s_i$ ($i=1, \ldots, n+1$). Then the measure can be expressed as
\be \int_0^\infty \prod_{r=1}^{n+1} d\s_r = \int_0^\infty d\s \s^n \Big[\int d\omega_n\Big], \ee
where the new variables are $\s$ defined by
\be \s = \sum_{r=1}^{n+1} \s_r,\ee
and the $n$ variables $\omega_i$ defined by 
\be \s \omega_i = \sum_{r=1}^i \s_r, \quad 1\leq i \leq n.\ee
Thus
\be 0 \leq \omega_1 \leq \ldots \leq \omega_n \leq 1.\ee
Also we have defined
\be \Big[\int d\omega_n\Big] \equiv \int_0^{1}d\omega_n \int_0^{\omega_n}d\omega_{n-1} \ldots \int_0^{\omega_3}d\omega_2 \int_0^{\omega_2}d\omega_1.\ee
The various loop momenta are always Euclidean in our analysis.

In the various diagrams below, the external momenta $k_i$ $(i=1,2,3,4)$ which satisfy $k_i^2=0$ flow into the diagrams. The Mandelstam variables are defined by $S = - G^{MN} (k_1 + k_2)_M (k_1 + k_2)_N, T = -G^{MN} (k_1 + k_4)_M (k_1 + k_4)_N, U = - G^{MN} (k_1 + k_3)_M (k_1 + k_3)_N$ where $G_{MN}$ is the M theory metric. The graviton momenta are non--vanishing only in the non--compact dimensions while the loop momenta are non--vanishing in all dimensions.  

\subsection{The one loop four graviton amplitude}

We first consider the four graviton amplitude at one loop, where we start with the expression for the amplitude in 11 dimensions, and then consider the expression in the compactified theory. In 11 uncompactified dimensions, the four graviton amplitude at one loop is given by
\be \mathcal{A}_{11}^{(1)} = \kappa_{11}^4[I(S,T) + I(S,U)+I(U,T)]\mathcal{R}^4,\ee
where
\be I(S,T) = \int \frac{d^{11}q}{(2\pi)^{11}q^2 (q+k_1)^2 (q+k_1 + k_2)^2 (q-k_4)^2} = I(T,S),\ee
and $2\kappa_{11}^2 =(2\pi)^8l_{11}^9$, where $l_{11}$ is the 11 dimensional Planck length.  This is depicted by figure 1.

\begin{figure}[ht]
\begin{center}
\[
\mbox{\begin{picture}(150,105)(0,0)
\includegraphics[scale=.7]{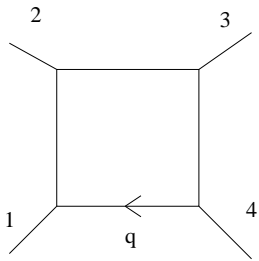}
\end{picture}}
\]
\caption{The one loop diagram $I(S,T)$ }
\end{center}
\end{figure}

Compactifying on $\mathbb{R}^{9-d,1} \times T^{d+1}$ where $\mathcal{V}_{d+1}$ is the dimensionless volume of $T^{d+1}$ in units of $l_{11}^{d+1}$ in the M theory metric, the $D$ dimensional amplitude is given by\footnote{We drop the subscript $D$ in $\mathcal{A}^{(1)}_D$ for brevity, which shall be the practice henceforth.}
\be \mathcal{A}^{(1)} = \kappa_{11}^2 \kappa_D^2 [I(S,T) + I(S,U)+I(U,T)]\mathcal{R}^4,\ee
where
\be \label{genD} I(S,T) = \sum_{m_I}\int \frac{d^D q}{(2\pi)^D (q^2 +  {\bf{m}}^2) ((q+k_1)^2 + {\bf{m}}^2)((q+k_1 + k_2)^2 + {\bf{m}}^2)((q-k_4)^2+ {\bf{m}}^2)} ,\ee
on using the relation
\be \frac{1}{\kappa_D^2} = \frac{l_{11}^{d+1}\mathcal{V}_{d+1}}{\kappa_{11}^2}.\ee
In \C{genD}, $m_I$ $(I=1,\ldots,d+1)$ are integers from the KK momenta and we have defined 
\be {\bf{m}}^2 \equiv G^{IJ} m_I m_J/l_{11}^2.\ee

We now evaluate \C{genD} by using the Schwinger representation of the propagators and performing the momentum integral. This gives us
\be \label{val1}(2\pi)^D I(S,T) = \pi^{D/2}\int_0^\infty d\s \s^{(d-4)/2}\Big[\int d\omega_3\Big]\sum_{m_I} e^{-\s {\bf{m}}^2 - \s Q(S,T;\omega_i)},\ee
where
\be Q(S,T;\omega_i) = -S\omega_1 (\omega_3 - \omega_2) -T(\omega_2-\omega_1)(1-\omega_3).\ee
$I(S,T)$ is defined for negative $S$ and $T$ to ensure convergence and then defined by analytic continuation elsewhere.

The $\mathcal{R}^4$ contribution is obtained by setting $m_I=0$ in \C{val1}, while the remaining contribution is given by  
\be \label{val2} \pi^{D/2}\int_0^\infty d\s \s^{(d-4)/2} \Big[\int d\omega_3\Big]\sum_{m_I} e^{-\s {\bf{m}}^2} \Big(e^{- \s Q(S,T;\omega_i)} - 1\Big).\ee
The $m_I=0$ term contributes in \C{val2} to the non--local part of the action. These contributions which are moduli independent are not relevant for our purposes. They yield terms power behaved or logarithmic in the momenta depending on the spacetime dimensions, where the logarithmic terms can be separately calculated to give moduli independent contributions finally leading to terms involving only ${\rm ln} g_d$ in the Einstein frame. We shall neglect such contributions in our analysis.

The remaining local contributions are given by
\be \label{S}\pi^{D/2}\sum_{m_I}' \sum_{n=1}^\infty\int_0^\infty d\s \s^{(d-4)/2+n} e^{-\s {\bf{m}}^2}\Big[\int d\omega_3\Big] \frac{(-  Q(S,T;\omega_i))^n}{n!} \ee
where we have excluded the term with $m_I=0$. The $n=1$ term in the sum vanishes using $S+T+U=0$, however we have kept it as it shall be useful for our purposes.   

Thus the $\mathcal{R}^4$ term is given by
\be \label{11}\mathcal{A}_{\mathcal{R}^4} = \frac{\pi^{D/2}\kappa_{11}^2 \kappa_D^2}{2(2\pi)^D} \mathcal{R}^4 \int_0^\infty d\s \s^{(d-4)/2} \sum_{m_I} e^{-\s {\bf{m}}^2}\ee
which receives no more contributions beyond one loop, while
the $D^4 \mathcal{R}^4$ term is given by
\be \label{12}\mathcal{A}^{(1)}_{D^4\mathcal{R}^4} =  \frac{\pi^{D/2}\kappa_{11}^2 \kappa_D^2}{2\cdot 6!(2\pi)^D} \s_2 \mathcal{R}^4 \sum'_{m_I} \int_0^\infty d\s \s^{d/2} e^{-\s {{\bf{m}}}^2},\ee
where we have defined  
\be \s_n \equiv S^n + T^n + U^n.\ee 
Now let us consider the $n=1$ term in \C{S} which gives us
\be \label{10}\mathcal{A}_{D^2\mathcal{R}^4} =  \frac{2\pi^{D/2}\kappa_{11}^2 \kappa_D^2}{5!(2\pi)^D} \s_1 \mathcal{R}^4 \sum'_{m_I} \int_0^\infty d\s \s^{(d-2)/2} e^{-\s {{\bf{m}}}^2}.\ee
We stress that since we are calculating S--matrix elements this vanishes, but we shall see later that the moduli dependent coefficient in \C{10} given by
\be \sum'_{m_I} \int_0^\infty d\s \s^{(d-2)/2} e^{-\s {{\bf{m}}}^2}\ee
arises as the coefficient function of a logarithmically divergent term that contributes to the $D^6\mathcal{R}^4$ amplitude. Hence we simply define this contribution to be coming from the one loop $D^2\mathcal{R}^4$ amplitude\footnote{Whether this receives contributions beyond one loop will not affect our analysis, as every expression in supergravity has to be replaced by its U--duality invariant completion.}.

\subsection{The two loop four graviton amplitude}

For the two and three loop amplitudes, we directly write down the expression for the four graviton amplitude in the theory compactified on $\mathbb{R}^{9-d,1} \times T^{d+1}$.

\begin{figure}[ht]
\begin{center}
\[
\mbox{\begin{picture}(310,90)(0,0)
\includegraphics[scale=.65]{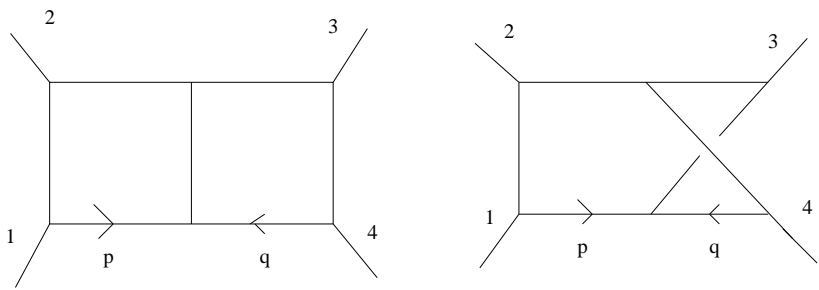}
\end{picture}}
\]
\caption{The two loop diagrams $I^P (S,T)$ and $I^{NP} (S,T)$}
\end{center}
\end{figure}
On compactifying on $\mathbb{R}^{9-d,1} \times T^{d+1}$ the four graviton amplitude at two loops is given by
\bea \mathcal{A}^{(2)} &=& \kappa_{11}^2 \kappa_D^4 \Big[S^2 \Big( I_P (S,T) + I_P (S,U) +I_{NP} (S,T) + I_{NP}(S,U)\Big) \non \\ &&+ T^2 \Big( I_P (T,S) + I_P (T,U) +I_{NP} (T,S) + I_{NP}(T,U) \Big) \non \\ &&+ U^2 \Big( I_P (U,S) + I_P (U,T) +I_{NP} (U,S) + I_{NP}(U,T)\Big)\Big]\mathcal{R}^4,\eea
where the planar contribution is given by
\bea \label{planar}I_P (S,T) = \sum_{m_I,n_I}\int \frac{d^D p}{(2\pi)^D} \int \frac{d^Dq}{(2\pi)^D} \frac{1}{(p^2 +{\bf{m}}^2)((p-k_1)^2 +{\bf{m}}^2)((p-k_1 - k_2)^2 +{\bf{m}}^2)}\non \\ \times \frac{1}{((p+q)^2+{\bf{(m+n)}}^2)(q^2+{\bf{n}}^2)( (q-k_4)^2+{\bf{n}}^2) ((q-k_3 - k_4)^2+{\bf{n}}^2)},\non \\ \eea
while the non--planar contribution is given by
\bea \label{nplanar}I_{NP} (S,T) = \sum_{m_I,n_I}\int \frac{d^D p}{(2\pi)^D} \int \frac{d^Dq}{(2\pi)^D} \frac{1}{(p^2+{\bf{m}}^2) ((p-k_1)^2+{\bf{m}}^2) ((p-k_1 - k_2)^2 +{\bf{m}}^2)}\non \\ \times \frac{1}{((p+q)^2 +{\bf{(m+n)}}^2)(q^2 +{\bf{n}}^2)((q-k_4)^2 +{\bf{m}}^2)((p+q+k_3)^2+{\bf{(m+n)}}^2)}\non \\ \eea
as depicted by figure 2, where the momenta denote the 11 dimensional loop momenta $p_M, q_M$ which are then split into the non--compact momenta $p_\mu, q_\m$ and the KK momenta denoted by the integers $m_I, n_I$ respectively. We shall also denote the loop momenta in the three loop diagrams in the same way later on.

Now for the $D^4\mathcal{R}^4$ amplitude, from \C{planar} we have that 
\be (2\pi)^{2D}I_P (0,0) = \frac{\pi^D}{4} \sum_{m_I, n_I} \int_0^\infty d\s d\l d\r \frac{\s^2\l^2}{\Delta_2^{(10-d)/2}(\s,\l,\r)} e^{-(\s {{\bf{m}}}^2 + \l {{\bf{n}}}^2 +\r{{\bf{(m+n)}}}^2)},\ee
on introducing Schwinger parameters and performing the momentum integrals, and
similarly
\be (2\pi)^{2D}I_{NP} (0,0) = \frac{\pi^D}{2} \sum_{m_I, n_I} \int_0^\infty d\s d\l d\r \frac{\l^2 \s\r}{\Delta_2^{(10-d)/2}(\s,\l,\r)} e^{-(\s {{\bf{m}}}^2 + \l{{\bf{n}}}^2 +\r{{\bf{(m+n)}}}^2)},\ee
from \C{nplanar},
where
\be \Delta_2 (\s,\l,\r) = \s\l +\l\r +\r\s.\ee
Thus adding these contributions, we have that
\bea \label{21}\mathcal{A}_{D^4\mathcal{R}^4}^{(2)} = \frac{\pi^D\kappa_{11}^2\kappa_D^4}{6(2\pi)^{2D}}\s_2\mathcal{R}^4  \sum_{m_I, n_I} \int_0^\infty \frac{d\s d\l d\r}{\Delta_2^{(6-d)/2}(\s,\l,\r)}  e^{-(\s {{\bf{m}}}^2 + \l {{\bf{n}}}^2 +\r{{\bf{(m+n)}}}^2)}. \eea
Expanding to the next order in the low momentum expansion, we get that
\bea \label{22}\mathcal{A}_{D^6\mathcal{R}^4}^{(2)} = \frac{\pi^D\kappa_{11}^2\kappa_D^4}{72(2\pi)^{2D}}\s_3\mathcal{R}^4  \sum_{m_I, n_I} \int_0^\infty \frac{d\s d\l d\r}{\Delta_2^{(6-d)/2}(\s,\l,\r)} \Big(\s+\l+\r- \frac{5\s\l\r}{\Delta_2(\s,\l,\r)}\Big) \non \\ \times e^{-(\s {{\bf{m}}}^2 + \l {{\bf{n}}}^2 +\r{{\bf{(m+n)}}}^2)}.\eea

Thus adding \C{12} and \C{21} we see that the $D^4\mathcal{R}^4$ amplitude is given by
\bea \label{D22}\mathcal{A}_{D^4\mathcal{R}^4} &=&  \frac{\pi^{D/2}\kappa_{11}^2\kappa_D^2}{6(2\pi)^D}\s_2\mathcal{R}^4  \Big[\frac{1}{240}\sum_{m_I}'\int_0^\infty \frac{d\s}{ \s^{-d/2}} e^{-\s {{\bf{m}}}^2}  \non \\ &&+\frac{\pi^{D/2}\kappa_D^2}{(2\pi)^D} \sum_{m_I,n_I}\int_0^\infty \frac{d\s d\l d\r }{\Delta_2^{(6-d)/2}(\s,\l,\r)}e^{-(\s {{\bf{m}}}^2 + \l {{\bf{n}}}^2 +\r{{\bf{(m+n)}}}^2)} \Big] ,\eea
which receives no more contributions beyond two loops.

\subsection{The three loop four graviton amplitude}

\begin{figure}[ht]
\begin{center}
\[
\mbox{\begin{picture}(400,80)(0,0)
\includegraphics[scale=.55]{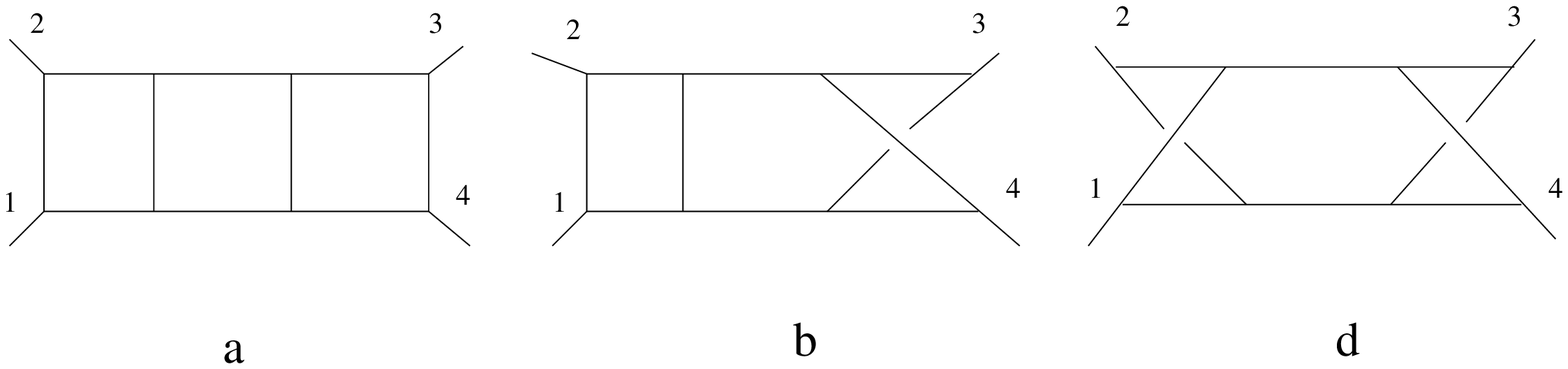}
\end{picture}}
\]
\caption{Three loop diagrams from the ladder skeleton}
\end{center}
\end{figure}

\begin{figure}[ht]
\begin{center}
\[
\mbox{\begin{picture}(350,190)(0,0)
\includegraphics[scale=.6]{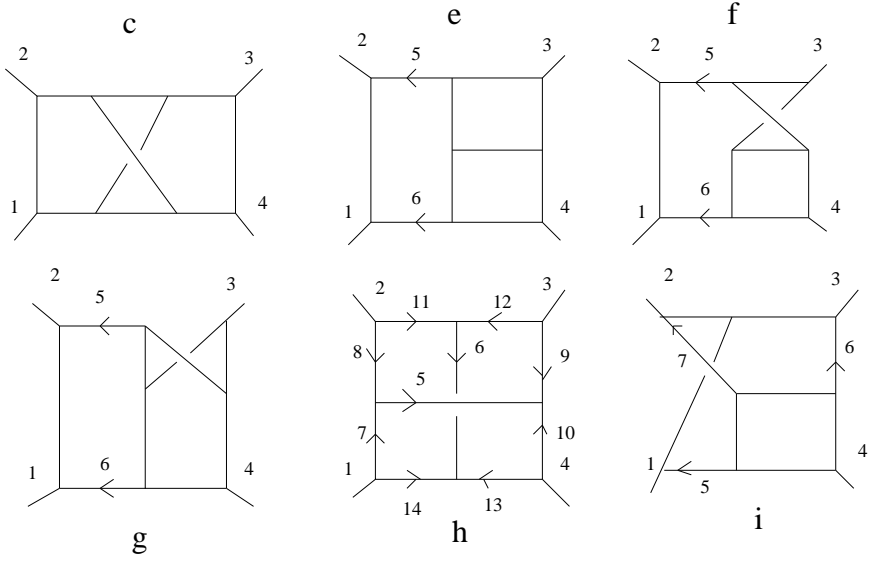}
\end{picture}}
\]
\caption{Three loop diagrams from the Mercedes skeleton}
\end{center}
\end{figure}

Now let us consider the four graviton amplitude at three loops. There are nine loop diagrams given by figures 3 and 4, which have been obtained  using unitarity cut techniques. The structure of the amplitude is more involved compared to the one and two loop amplitudes. Unlike the lower loop amplitudes which have only one underlying skeleton diagram, the three loop amplitude has two underlying skeleton diagrams--the ladder and Mercedes skeleton diagrams.

Of these diagrams, the diagrams $a,b$ and $d$ are obtained from the ladder skeleton diagram, while the rest are obtained from the Mercedes skeleton diagram. While the integrands for the loop diagrams $a,b,c$ and $d$ have numerator 1, the numerators in the integrands for the other loop diagrams have non--trivial dependence on the loop momenta as well as on the external momenta, which are given below. Hence unlike the lower loop amplitudes, the three loop amplitude is not simply given by massless $\varphi^3$ scalar field theory. 

The three loop amplitude is given by
\bea \label{totcont}
\mathcal{A}^{(3)} &=& \kappa_{11}^2\kappa_D^6\sum_{S_3} \Big[ I^{(a)} + I^{(b)} + \frac{1}{2} I^{(c)} + \frac{1}{4} I^{(d)} + 2 I^{(e)} + 2 I^{(f)} + 4 I^{(g)} + \frac{1}{2} I^{(h)} + 2 I^{(i)}\Big] \mathcal{R}^4 \non \\ &\equiv& \kappa_{11}^2\kappa_D^6 I_3 \mathcal{R}^4. \eea
where $S_3$ represents the 6 independent permutations of the external legs marked $\{1,2,3\}$ keeping the external leg $\{4\}$ fixed.

Now the numerators $N^{(x)}$ for the various integrands in the loop diagrams are given by~\cite{Bern:2008pv}
\bea \label{num}
N^{(a)} &=& N^{(b)} = N^{(c)} = N^{(d)} = S^4 , \non \\ N^{(e)} &=& N^{(f)} = N^{(g)} = S^2 \tau_{35} \tau_{46}, \non \\ N^{(h)} &=& \Big(S(\tau_{26} +\tau_{36}) +T(\tau_{15} +\tau_{25}) +ST\Big)^2 \non \\ &&+ \Big(S^2 (\tau_{26} +\tau_{36}) - T^2 (\tau_{15} +\tau_{25}) \Big) \Big(\tau_{17} +\tau_{28} +\tau_{39} +\tau_{4,10} \Big) \non \\&& +S^2 (\tau_{17}  \tau_{28} +\tau_{39} \tau_{4,10}) +T^2 (\tau_{28} \tau_{39}+ \tau_{17} \tau_{4,10}) + U^2 (\tau_{17} \tau_{39} + \tau_{28} \tau_{4,10}), \non \\ N^{(i)} &=& (S\tau_{45} - T\tau_{46})^2 -\tau_{27} (S^2 \tau_{45} + T^2 \tau_{46}) - \tau_{15} (S^2 \tau_{47}+ U^2 \tau_{46})\non \\ &&-  \tau_{36} (T^2 \tau_{47} +U^2 \tau_{45}) - l_5^2 S^2 T - l_6^2 S T^2 +\frac{l_7^2}{3}STU,\eea
where
\be \tau_{ij} = -2 k_i \cdot l_j ~(i \leq 4, j \geq 5).\ee
The momenta $l_i$ are denoted in figure 4. Thus on compactifying on $\mathbb{R}^{9-d,1} \times T^{d+1}$, each term in the expression \C{totcont} is of the form
\be \sum_{l_I,m_I,n_I}\int \frac{d^Dp}{(2\pi)^D} \int \frac{d^D q}{(2\pi)^D} \int \frac{d^Dr}{(2\pi)^D}\frac{N^{(x)}}{\mathcal{D}^{(x)}},\ee
where the denominator $\mathcal{D}^{(x)}$ is simply given by the product of massless propagators as in the one and two loop cases depending on the momentum labels in $x$.

Thus for the $D^6\mathcal{R}^4$ amplitude, only the diagrams $e,f,g,h$ and $i$ contribute leading to
\be \label{3}\mathcal{A}^{(3)}_{D^6\mathcal{R}^4} = \frac{5\pi^{3D/2}\kappa_{11}^2\kappa_D^6}{6(2\pi)^{3D}}\s_3\mathcal{R}^4\int_0^\infty d\Upsilon \Delta_3^{1-D/2} (\s,\lambda,\mu,\rho,\nu,\theta) F_L (\s,\lambda, \mu,\rho, \nu,\theta).\ee
In \C{3}, the measure factor is given by
\be  d\Upsilon \equiv d\s d\lambda d\mu d\rho d\nu d\theta,\ee
while the factor obtained by integrating over the loop momenta is given by
\bea \label{defD}\Delta_3 (\s,\lambda,\mu,\rho,\nu,\theta) &=& \s \lambda \mu +\rho\nu\theta + \s\mu (\rho+\nu+\theta) +\lambda\mu (\rho+\theta) +\s\lambda (\nu+\theta)\non \\ &&+ \mu\nu(\rho+\theta) +\s\rho(\nu+\theta) +\lambda(\rho\nu+\nu\theta+\rho\theta).\eea
Finally the lattice factor is given by
\bea \label{defF} &&F_L (\s,\lambda,\mu,\rho,\nu,\theta)  = \sum_{l_I, m_I, n_I} e^{-\Big( \n {{\bf{l}}}^2 +\m {{\bf{m}}}^2 +\r {{\bf{n}}}^2 +\l {{\bf{(l+m)}}}^2 +\theta {{\bf{(m+n)}}}^2 +\s{{\bf{(l+m+n)}}}^2 \Big)/l_{11}^2}. \eea

Thus the expressions \C{11} and \C{D22} give the complete $\mathcal{R}^4$ and $D^4\mathcal{R}^4$ amplitudes in maximal supergravity, while \C{10} defines the $D^2\mathcal{R}^4$ amplitude. Also \C{22}, \C{3} along with the one loop amplitude give the complete $D^6\mathcal{R}^4$ amplitude. These will play a central role in our analysis.

\section{Logarithmically divergent contributions: preliminaries}

Before proceeding with the detailed calculations, let us consider some simple cases and some generalities to understand the primary logic. Consider the one loop $\mathcal{R}^4$ amplitude which involves the integral (see appendix A)
\bea \label{many}\sum_{m_I} \int \frac{d^D q}{(q^2 + {{\bf{m}}}^2)^4} &=& \frac{\pi^{D/2}}{6}  \sum_{m_I} \int_{\Lambda^{-2}}^\infty d\s \s^{(d-4)/2} e^{-\s {{\bf{m}}}^2 } \non \\ &=& \frac{\pi^{11/2}}{6}\sum_{\hat{m}^I} \int_0^{\Lambda^2} d\hat\s \sqrt{\hat\s} e^{-\pi^2 l_{11}^2 G_{IJ} \hat{m}^I \hat{m}^J \hat\s}. \eea
Let us focus on the ultraviolet nature of the various integrals, which is our primary focus. The first expression in \C{many} has a logarithmic ultraviolet divergence in 8 dimensions. The KK modes are irrelevant for these large values of loop momenta. In fact, the amplitude in supergravity (which is moduli independent) is obtained by setting the KK momenta to zero, which has the UV structure as mentioned above, which is nothing but the process of dimensional reduction. This is the kind of divergence we are interested in, which we shall refer to as the field theory divergence, which has no moduli dependence. The second expression in \C{many} is in the Schwinger representation, where the UV divergence arises from $\s\rightarrow 0$. Of course, the divergence is logarithmic in 8 dimensions arising from the
\be \int_{\Lambda^{-2}}^\infty \frac{d\s}{\s}\ee     
integral, which reproduces the field theory result. Again the KK modes play no role in the analysis. We shall find the Schwinger representation of the amplitude most useful in our analysis as this involves manipulating lesser number of terms than using the Feynman propagators to do the integrals. This is because when we expand the integrals using Feynman propagators at various orders in the momentum expansion, there are many terms whereas all these are obtained from simply expanding the exponential in the Schwinger representation to the required order in the momentum expansion. The third expression in \C{many} is obtained by Poisson resumming the second expression, which has a leading UV divergence as $\hat{\s} \rightarrow \infty$. This is contained in the sector when all $\hat{m}^I=0$, and is of the form $\Lambda^3$ which is the primitive UV divergence of the 11 dimensional theory. This is not the field theory divergence of the compactified theory as Poisson resummation which transforms from the KK mode basis to the winding mode basis essentially reorganizes the divergence and the zero winding sector reproduces the UV structure of the parent theory, independent of the details of the compactification. These arguments are true in general. Thus we shall refer to the logarithmic UV divergence as a field theory logarithmic divergence which arises from simply setting the KK modes to zero in specific dimensions.            

This is the general structure we shall follow in our analysis. In order to isolate the moduli independent field theory logarithmic divergence from certain loops we shall simply set the KK momenta to zero for the corresponding loop integrals and obtain the divergent contribution on integrating over the loop momenta. The KK modes for the remaining loops in the amplitude will be summed over all integers. On integrating over these loop momenta, this leads to moduli dependent coefficient functions which in the complete amplitude, multiply the field theory logarithmic divergences obtained from the other loops. The moduli dependent part will produce $SL(d+1,\mathbb{Z})$ invariant modular forms for compactifications on $T^{d+1}$\footnote{Note that their field theory divergences which result from setting the KK momenta to zero are very different from that obtained by keeping the KK modes and performing Poisson resummation as discussed above.}. This is where the structure crucially differs from ordinary supergravity amplitudes, where the KK momenta are all set to zero and hence all divergences are moduli independent. On the other hand, our calculations will produce moduli independent divergences from some loops like in ordinary supergravity, but with moduli dependent coefficient functions resulting from the infinite sum over all KK momenta in the other loops. This partially captures the stringy nature of the amplitude.

To illustrate this point we now consider the systematics of a simple two loop amplitude, which we shall later generalize. To be specific, consider the planar diagram contribution to the two loop $D^4\mathcal{R}^4$ amplitude which involves the integral
\be \label{I}\sum_{m_I,n_I}\int \frac{d^Dp d^Dq}{(p^2 +{\bf{m}}^2)^3 (q^2+{\bf{n}}^2)^3((p+q)^2+{\bf{(m+n)}}^2)},\ee  
as depicted by the diagram on the left in figure 2. There is a one loop field theory nested logarithmic UV divergence associated with this integral. To obtain this, we first set $n_I =0$ so that the $q$ momentum loop integral is dimensionally reduced. Next we perform a simple binomial expansion for the propagator involving $(p+q)^2+{\bf{m}}^2$ leading to
\be \label{exP}\frac{1}{(p+q)^2+{\bf{m}}^2} = \frac{1}{p^2+{\bf{m}}^2} \Big( 1- \frac{q^2 + 2p\cdot q}{p^2 + {\bf{m}}^2}+\ldots \Big)\ee
Thus from the structure of the propagators that result from this expansion we see that every individual term can be expressed as a product of one loop integrals.   

Let us consider the contribution coming from the first term in \C{exP}. The integral \C{I} factorizes into the moduli independent $q$ loop integral, and the moduli dependent $p$ loop integral, and is given by
\be \label{I2}\mathcal{I} = \sum_{m_I}\int \frac{d^Dp}{(p^2 +{\bf{m}}^2)^4} \int \frac{d^D q}{(q^2)^3} ,\ee 
which has a logarithmic UV divergence in 6 dimensions, with a moduli dependent coefficient function involving the one loop $\mathcal{R}^4$ amplitude\footnote{There is an identical contribution from the $p$ loop integral when $m_I =0$ and we keep the term $1/(q^2 + {\bf{n}}^2)$ in the binomial expansion for $1/((p+q)^2 +{\bf{n}}^2)$, and hence the total contribution is twice the expression in \C{I2}.}.

One can now keep more terms in \C{exP} and generalize the analysis. Note that effectively this amounts to taking $\vert p \vert >> \vert q \vert$, in which to leading order
\be (p+q)^2+{\bf{m}}^2 \rightarrow p^2 + {\bf{m}}^2  \ee  
which reproduces what we have above, while keeping the other terms in \C{exP} amounts to a perturbative expansion in large $\vert p \vert$. Hence we shall simply refer to this as the large $\vert p \vert$ expansion, keeping in mind that the loop momenta have to be integrated over all values at the end.

Now let us consider \C{I2} in some detail. 
 Expressed in terms of Schwinger parameters, we have that      
\be \label{match}\mathcal{I} =  \frac{\pi^D}{12}\int_0^\infty d\l \l^{(d-6)/2} \sum_{m_I}\int_0^\infty d\s \s^{(d-4)/2} e^{-\s {\bf{m}}^2} .\ee

Now let us express the integral \C{I} directly in the Schwinger representation, which gives us
\be \frac{\pi^D}{4} \sum_{m_I, n_I} \int_0^\infty d\s d\l d\r \frac{\s^2\l^2}{(\s\l+\l\r+\r\s)^{D/2}} e^{-(\s {{\bf{m}}}^2 + \l {{\bf{n}}}^2 +\r{{\bf{(m+n)}}}^2)}.\ee
Here $\s,\l,\r$ are the Schwinger parameters corresponding to the loop momenta $p_\m, q_\mu, (p+q)_\mu$ as well as the KK momenta $m_I,n_I,(m+n)_I$ respectively.  Now let us consider the leading contribution in the limit discussed above. First setting $n_I=0$ gives us 
\be \label{I3}\frac{\pi^D}{4} \sum_{m_I} \int_0^\infty d\s d\l d\r \frac{\s^2\l^{(d-6)/2}}{(\s+\r)^{D/2}\Big[1+\s\r/\l(\s+\r)\Big]^{D/2}} e^{-(\s+\r) {{\bf{m}}}^2 }.\ee
In the large $\vert p \vert$ expansion, it is easy to check by directly performing the Gaussian integrals in $\vert p \vert$ and $\vert q \vert$ that the $\l(\s+\r)$ contribution dominates over the $\s\r$ contribution in $\l(\s+\r)+\r\s$ in this limit. Hence we obtain a perturbative expansion in the dimensionless parameter $\s\r/\l(\s+\r)$ in this regime of the loop momenta. We simply refer to it as the large $\l$ expansion (alternatively this is the small $\s,\r$ expansion)\footnote{The contribution from large $\vert q \vert$ thus corresponds to the large $\s$ expansion, which is equal to the one we calculate.}. Thus from \C{I3} the leading contribution is given by 
\be \label{I4}\frac{\pi^D}{4} \int_0^\infty d\l \l^{(d-6)/2} \sum_{m_I} \int_0^\infty d\s d\r \frac{\s^2}{(\s+\r)^{D/2}} e^{-(\s+\r) {{\bf{m}}}^2 }.\ee
Now we define
\be \label{simp}\m = \s+\r, \quad \omega = \s/\m\ee
such that
\be 0 \leq \omega \leq 1.\ee
Then we have that
\be \int_0^\infty d\s d\r \frac{\s^2}{(\s+\r)^{D/2}} e^{-(\s+\r) {{\bf{m}}}^2 } = \frac{1}{3}\int_0^\infty d\m \mu^{(d-4)/2} e^{-\m {{\bf{m}}}^2}, \ee
and hence \C{I4} is exactly equal to $\mathcal{I}$ in \C{match}. Thus we have isolated a particular logarithmic divergence in the Schwinger representation and related it to  the same divergence obtained using Feynman propagators. This is the procedure we shall generally follow in our analysis.

Hence we see the origin of such logarithmically divergent terms in these simple examples, which generalizes to the cases we are interested in. Now we generalize the analysis for the complete amplitudes given the explicit expressions for the supergravity amplitudes. We perform the analysis in the Schwinger representation of the amplitudes as this considerably simplifies our calculations.

\section{Logarithmically divergent contributions: the detailed analysis}

We consider toroidal compactifications to $D$ dimensions where $3 \leq D \leq 10$. In fact, since we are looking at S--matrix elements the four graviton amplitude trivially vanishes for $D=3$ as the $\mathcal{R}^4$ term only involves the Weyl tensor. Hence we focus only on $4 \leq D \leq 10$. We are only interested in the contributions to the $\mathcal{R}^4$, $D^4\mathcal{R}^4$ and $D^6\mathcal{R}^4$ amplitudes where the logarithmic UV divergence has moduli dependent coefficient functions. Hence we need to consider contributions from two and three loops only.

\subsection{The contributions from two loops}

First we consider the contribution to the two loop $D^4\mathcal{R}^4$ amplitude from \C{21}. This expression has the symmetry under interchange of the three Schwinger parameters which represent the underlying two loop skeleton diagram, depicted in figure 5.  

\begin{figure}[ht]
\begin{center}
\[
\mbox{\begin{picture}(120,90)(0,0)
\includegraphics[scale=.5]{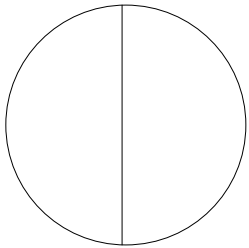}
\end{picture}}
\]
\caption{The two loop ladder skeleton}
\end{center}
\end{figure}

To obtain the logarithmic contributions, we set $n_I=0$ and perform the large $\l$ expansion. There is an overall factor of 3 from the possibility of choosing the 3 Schwinger parameters, alternatively the 3 links of the two loop skeleton diagram. Thus, we have that 
\bea \mathcal{A}_{D^4\mathcal{R}^4}^{(2)} &=& \frac{\pi^D\kappa_{11}^2\kappa_D^4}{2(2\pi)^{2D}}\s_2\mathcal{R}^4  \sum_{m_I} \int_0^\infty \frac{d\s d\l d\r}{[\l(\s+\r)]^{(6-d)/2}}  e^{-(\s+\r) {{\bf{m}}}^2 }\Big[ 1-\frac{(6-d)\s\r}{2\l(\s+\r)} + O(1/\lambda^2)\Big] \non \\ &=& \frac{\pi^D\kappa_{11}^2\kappa_D^4}{2(2\pi)^{2D}}\s_2\mathcal{R}^4   \sum_{m_I} \Big[ \int_0^\infty \frac{d\l}{\l^{(6-d)/2}} \int_0^\infty \frac{d\m}{ \mu^{(4-d)/2}} e^{-\m {{\bf{m}}}^2}  \non \\ &&+ \frac{d-6}{12} \int_0^\infty \frac{d\l}{\l^{(8-d)/2}} \int_0^\infty \frac{d\m}{ \mu^{(2-d)/2}} e^{-\m {{\bf{m}}}^2}\Big].  \eea
In the first line we have performed the large $\l$ expansion, where we have kept the first two terms only. In the second line we have used \C{simp} to simplify these two contributions. The first term yields a logarithmic divergence in 6 dimensions, with  the coefficient involving the one loop $\mathcal{R}^4$ amplitude. The second yields the same in 4 dimensions (thus $d=6$), with a coefficient proportional to the one loop $D^2\mathcal{R}^4$ amplitude which vanishes onshell. However, the overall coefficient is proportional to $d-6$ and hence vanishes. The terms of $O(1/\l^2)$ which we have ignored in the first line do not produce any more logarithmic divergences for $4 \leq D \leq 10$. Hence the total logarithmic divergence is given by
\be \label{log1} \mathcal{A}_{D^4\mathcal{R}^4}^{(2)} = \frac{\pi^D\kappa_{11}^2\kappa_D^4}{2(2\pi)^{2D}}\s_2\mathcal{R}^4   \delta_{D,6} \int_0^\infty \frac{d\l}{\l^{(6-d)/2}} \sum_{m_I}\int_0^\infty \frac{d\m}{ \mu^{(4-d)/2}} e^{-\m {{\bf{m}}}^2} .\ee     

Next we consider the contribution to the two loop $D^6\mathcal{R}^4$ amplitude in \C{22}. Proceeding as above, the relevant terms are given by 
\be \label{more}\mathcal{A}_{D^6\mathcal{R}^4}^{(2)} = \frac{\pi^D\kappa_{11}^2\kappa_D^4}{24(2\pi)^{2D}}\s_3\mathcal{R}^4  \sum_{m_I} \int_0^\infty \frac{d\s d\l d\r}{[\l(\s+\r)]^{(6-d)/2}}  e^{-(\s+\r) {{\bf{m}}}^2}\Big[ \l + f_1 (\s,\r) +\frac{f_2 (\s,\r)}{\l}+O(1/\l^2)\Big],\ee
where
\bea f_1 (\s,\r) &=& \s+\r - \frac{(16-d)\s\r}{2(\s+\r)}, \non \\ f_2 (\s,\r) &=& \frac{5(\s\r)^2}{(\s+\r)^2}+(6-d)\Big[\frac{(28-d)(\s\r)^2}{8(\s+\r)^2} -\frac{\s\r}{2}\Big].\eea
The three terms in \C{more} have logarithmic divergences in 8, 6 and 4 dimensions respectively, while the terms we have ignored in \C{more} do not yield any more such divergences for $4 \leq D \leq 10$. Thus these contributions are given by
\bea \label{ymore}\mathcal{A}_{D^6\mathcal{R}^4}^{(2)} &=& \frac{\pi^D\kappa_{11}^2\kappa_D^4}{24(2\pi)^{2D}}\s_3\mathcal{R}^4  \sum_{m_I}  \Big[  \delta_{D,8} \int_0^\infty \frac{d\l}{\l^{(4-d)/2}} \int_0^\infty \frac{d\s d\r}{(\s+\r)^{(6-d)/2}} \non \\ &&+  \delta_{D,6} \int_0^\infty \frac{d\l}{\l^{(6-d)/2}} \int_0^\infty \frac{d\s d\r}{(\s+\r)^{(6-d)/2}}\Big(\s+\r - \frac{6\s\r}{\s+\r}\Big)\non \\ &&+  5\delta_{D,4} \int_0^\infty \frac{d\l}{\l^{(8-d)/2}}\int_0^\infty \frac{d\s d\r (\s\r)^2}{(\s+\r)^{(10-d)/2}} \Big]e^{-(\s+\r) {{\bf{m}}}^2}.\eea 
We now use \C{simp} to simplify the $\s,\r$ integrals. The second term in \C{ymore} yields a contribution proportional to the one loop $D^2\mathcal{R}^4$ amplitude. However the $\omega$ integral yields
\be \int_0^1 d\omega \Big[1-6\omega(1-\omega)\Big]=0,\ee
and hence this contribution vanishes. Thus we get that
\bea \label{log2}\mathcal{A}_{D^6\mathcal{R}^4}^{(2)} &=& \frac{\pi^D\kappa_{11}^2\kappa_D^4}{24(2\pi)^{2D}}\s_3\mathcal{R}^4 \Big[  \delta_{D,8} \int_0^\infty \frac{d\l}{\l^{(4-d)/2}} \sum_{m_I}\int_0^\infty \frac{d\m}{ \mu^{(4-d)/2}} e^{-\m {{\bf{m}}}^2} \non \\ &&+  \frac{\delta_{D,4}}{6} \int_0^\infty \frac{d\l}{\l^{(8-d)/2}}\sum_{m_I}'\int_0^\infty \frac{d\m}{ \mu^{-d/2}} e^{-\m {{\bf{m}}}^2} \Big].\eea 
Note that the moduli dependent contributions involve the one loop $\mathcal{R}^4$ and $D^4\mathcal{R}^4$ amplitudes in 8 and 4 dimensions respectively\footnote{We have restricted the sum to $m_I \neq 0$ in the second term to obtain the $D^4\mathcal{R}^4$ one loop amplitude as explained before.}.

\subsection{The contributions from three loops}

We now consider the contribution from the three loop $D^6\mathcal{R}^4$ amplitude in \C{3}. The parametrization of the Mercedes skeleton diagram is given in figure 6 depicting Schwinger parameters corresponding to KK momenta, hence the lattice sum is given by \C{defF}. This has the structure of a tetrahedron, with 6 edges and 4 faces.  

\begin{figure}[ht]
\begin{center}
\[
\mbox{\begin{picture}(300,110)(0,0)
\includegraphics[scale=.5]{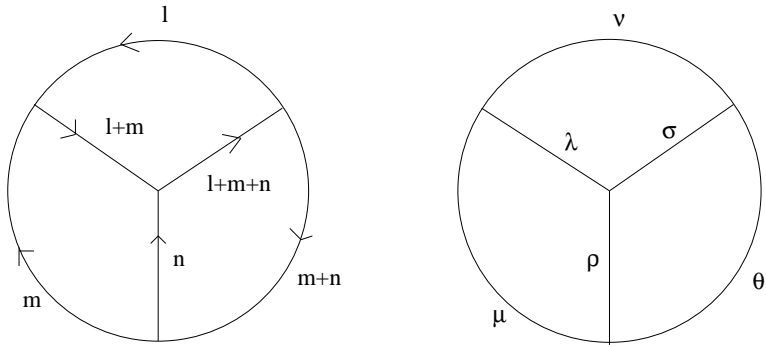}
\end{picture}}
\]
\caption{Parametrizing the Mercedes skeleton}
\end{center}
\end{figure}

First we consider the one loop logarithmic field theory divergences. To do so, we set $l_I =0$ in \C{3}, and make a large $\n$ expansion in $\Delta_3 (\s,\lambda,\mu,\rho,\nu,\theta)$. There are 6 such contributions coming from the 6 choices of the Schwinger parameters, alternatively from the 6 edges of the tetrahedron. This gives us   
\bea \label{A}&&\mathcal{A}^{(3)}_{D^6\mathcal{R}^4} = \frac{5\pi^{3D/2}\kappa_{11}^2\kappa_D^6}{(2\pi)^{3D}}\s_3\mathcal{R}^4\sum_{m_I,n_I}\int_0^\infty d\Upsilon e^{-((\m+\l) {{\bf{m}}}^2 + \r {{\bf{n}}}^2 +(\s+\theta){{\bf{(m+n)}}}^2)} \non \\ &&\times \Big[ \frac{1}{\n^{(8-d)/2}[\r(\s+\theta) + \r(\m+\l)+ (\s+\theta)(\m+\l)]^{(8-d)/2}}+O(1/\nu^{(10-d)/2})\Big].\eea
The structure of the resulting two loop diagram characterized by the Schwinger parameters $\m,\l,\r,\s,\theta$ is depicted in figure 7. This allows for a clear interpretation of the integrand in \C{A} diagrammatically. The terms we have ignored in \C{A} do not yield logarithmic terms in $4 \leq D \leq 10$.

\begin{figure}[ht]
\begin{center}
\[
\mbox{\begin{picture}(300,110)(0,0)
\includegraphics[scale=.5]{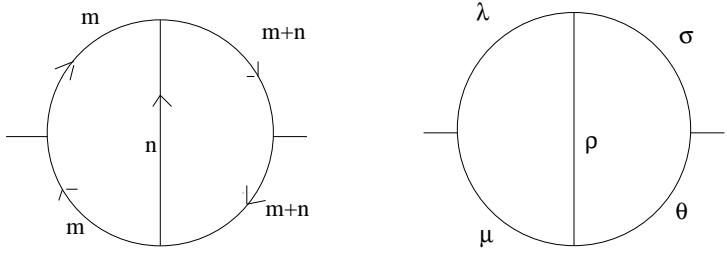}
\end{picture}}
\]
\caption{The underlying two loop diagram}
\end{center}
\end{figure}

The first term in \C{A} yields a logarithmic divergence in 4 dimensions, given by
\bea \label{A1}&&\mathcal{A}^{(3)}_{D^6\mathcal{R}^4} = \frac{5\pi^{3D/2}\kappa_{11}^2\kappa_D^6}{(2\pi)^{3D}}\s_3\mathcal{R}^4\delta_{D,4}\int_0^\infty \frac{d\n}{\n^{(8-d)/2}}\non \\ &&\times \sum_{m_I,n_I}\int_0^\infty \frac{d\s d\l d\r d\m \theta e^{-((\m+\l) {{\bf{m}}}^2 + \r {{\bf{n}}}^2 +(\s+\theta){{\bf{(m+n)}}}^2)}}{[\r(\s+\theta) + \r(\m+\l)+ (\s+\theta)(\m+\l)]^{(8-d)/2}}.\eea

To simplify \C{A1}, we define 
\be \a = \s+\theta, \quad \b = \m+\l, \quad \omega = \frac{\s}{\a}, \quad u = \frac{\l}{\b},\ee
thus
\be 0 \leq \omega, u \leq 1,\ee
leading to
\be \label{log3}\mathcal{A}^{(3)}_{D^6\mathcal{R}^4} = \frac{5\pi^{3D/2}\kappa_{11}^2\kappa_D^6}{3(2\pi)^{3D}}\s_3\mathcal{R}^4\delta_{D,4}\int_0^\infty \frac{d\n}{\n^{(8-d)/2}}\sum_{m_I,n_I}\int_0^\infty \frac{d\a d\b d\r }{\Delta_2^{(6-d)/2}(\a,\b,\r)}e^{-(\b {{\bf{m}}}^2 + \r {{\bf{n}}}^2 +\a{{\bf{(m+n)}}}^2)}.\ee
Thus the moduli dependent contribution involves the two loop $D^4\mathcal{R}^4$ amplitude.

Now let us consider the two loop logarithmic field theory divergences. To do so, we set $m_I = n_I = 0$ in \C{defF}, and make a large $\m,\r,\theta$ expansion in  $\Delta_3 (\s,\lambda,\mu,\rho,\nu,\theta)$. The remaining Schwinger parameters in the Mercedes skeleton are $\l,\s,\n$ which we refer to as dual Schwinger parameters, which parametrizes a face of the tetrahedron in figure 6. Thus there are 4 such contributions coming from the 4 choices of dual Schwinger parameters $(\l,\s,\n),(\l,\r,\m),(\r,\s,\theta),(\m,\n,\theta)$, alternatively from the 4 faces of the tetrahedron. Thus we get that
\bea \label{2t}\mathcal{A}^{(3)}_{D^6\mathcal{R}^4} = \frac{10\pi^{3D/2}\kappa_{11}^2\kappa_D^6}{3(2\pi)^{3D}}\s_3\mathcal{R}^4\sum_{l_I}\int_0^\infty \frac{d\Upsilon e^{-(\s+\l+\n){\bf{l}}^2}}{[(\s+\l+\n)\Delta_2 (\m,\r,\theta)]^{(8-d)/2}} \non \\ \times\Big[1+ \frac{(d-8)[\m\s(\l+\n) + \theta\l(\s+\n) +\r\n(\l+\s)]}{2(\s+\l+\n)\Delta_2 (\m,\r,\theta)}+O(1/\m^2)\Big].\eea
where the contributions beyond the first two terms can be ignored as they do not contribute logarithmic divergences for $4 \leq D \leq 10$. For the last term we have taken $\m,\r,\theta$ to be large and of the same order for the estimate (thus the second term is $O(1/\m)$).   

Let us focus on the first term in \C{2t} which yields a logarithmic divergence in 5 dimensions given by 
\be \label{t1}\mathcal{A}^{(3)}_{D^6\mathcal{R}^4} = \frac{10\pi^{3D/2}\kappa_{11}^2\kappa_D^6}{3(2\pi)^{3D}}\s_3\mathcal{R}^4 \delta_{D,5}\int_0^\infty \frac{d\m d\r d\theta}{\Delta_2^{(8-d)/2} (\m,\r,\theta)}\sum_{l_I}\int_0^\infty \frac{d\s d\l d\n}{(\s+\l+\n)^{(8-d)/2}} e^{-(\s+\l+\n){\bf{l}}^2}.\ee
Defining 
\be \label{range}\a = \s+\l+\n, \quad \omega_1 = \frac{\s}{\a}, \quad \omega_2 = \frac{\s+\l}{\a} \ee
such that
\be 0 \leq \omega_1 \leq \omega_2 \leq 1,\ee
the moduli dependent integral can be simplified, and \C{t1} gives 
\be \label{log4}\mathcal{A}^{(3)}_{D^6\mathcal{R}^4} = \frac{5\pi^{3D/2}\kappa_{11}^2\kappa_D^6}{3(2\pi)^{3D}}\s_3\mathcal{R}^4 \delta_{D,5}\int_0^\infty \frac{d\m d\r d\theta}{\Delta_2^{(8-d)/2} (\m,\r,\theta)}\sum_{l_I}\int_0^\infty \frac{d\a}{\a^{(4-d)/2}} e^{-\a{\bf{l}}^2}.\ee
Thus the moduli dependent term involves the one loop $\mathcal{R}^4$ amplitude. 

Next consider the second term in \C{2t} which yields a logarithmic divergence in 4 dimensions. Using \C{range}, we simplify the moduli dependent term to get
\bea \mathcal{A}^{(3)}_{D^6\mathcal{R}^4} = -\frac{5\pi^{3D/2}\kappa_{11}^2\kappa_D^6}{18(2\pi)^{3D}}\s_3\mathcal{R}^4\delta_{D,4}\int_0^\infty d\m d\r d\theta \frac{(\m+\r+\theta)}{\Delta_2^{(10-d)/2} (\m,\r,\theta)}\sum_{l_I}\int_0^\infty \frac{d\a}{\a^{(2-d)/2}} e^{-\a{\bf{l}}^2}. \non \\ \eea
The moduli independent term also simplifies as it is a total derivative giving us
\bea \label{D}(d-8)\int_0^\infty d\m d\r d\theta \frac{(\m+\r+\theta)}{\Delta_2^{(10-d)/2} (\m,\r,\theta)} &=& \int_0^\infty d\m d\r d\theta \Big(\frac{\p}{\p\m} +\frac{\p}{\p\r}+\frac{\p}{\p\theta}\Big)\frac{1}{\Delta_2^{(8-d)/2} (\m,\r,\theta)}\non \\ &=& -3 \int_0^\infty \frac{d\r}{\r^{(8-d)/2}} \int_0^\infty \frac{d\theta}{\theta^{(8-d)/2}},\eea
leading to
\bea \label{log5}\mathcal{A}^{(3)}_{D^6\mathcal{R}^4} = -\frac{5\pi^{3D/2}\kappa_{11}^2\kappa_D^6}{12(2\pi)^{3D}}\s_3\mathcal{R}^4\delta_{D,4}\int_0^\infty \frac{d\r}{\r^{(8-d)/2}} \int_0^\infty \frac{d\theta}{\theta^{(8-d)/2}}\sum_{l_I}'\int_0^\infty \frac{d\a}{\a^{(2-d)/2}} e^{-\a{\bf{l}}^2}.  \eea

Thus the moduli dependent part involves the coefficient function of the one loop $D^2\mathcal{R}^4$ amplitude\footnote{We have restricted to the $l_I \neq 0$ sum to relate to the one loop $D^2\mathcal{R}^4$ amplitude.}. 
Unlike the earlier cases where this contribution vanishes, the coefficient of this contribution is non--vanishing. 

Note that expressing the various amplitudes in the Schwinger representation simplifies calculations considerably. The various loop diagrams add up to give simple expressions which can them be manipulated to give the divergent contributions.

\subsection{The logarithmically divergent contributions}

Thus we have obtained the complete expression for the logarithmic ultraviolet divergences to the $D^4\mathcal{R}^4$ and $D^6\mathcal{R}^4$ amplitudes which are non--vanishing only in certain specific dimensions. The moduli dependent coefficient functions that multiply these logarithms are also completely determined by the structure of the multiloop supergravity amplitudes, and involve the $\mathcal{R}^4$, $D^2\mathcal{R}^4$ and $D^4\mathcal{R}^4$ amplitudes.

We would now like to analyze these logarithmic UV divergences using dimensional regularization. Hence we express the various integrals over Schwinger parameters in terms of momentum integrals which we then evaluate using dimensional regularization and isolate the pole terms in $\epsilon$. To do so, we use the identities
\bea &&\frac{\pi^{D/2}}{2} \int_0^\infty \frac{d\l}{\l^{(6-d)/2}} = \int \frac{d^D q}{(q^2)^3}, \quad \frac{\pi^{D/2}}{6} \int_0^\infty \frac{d\l}{\l^{(4-d)/2}} = \int \frac{d^D q}{(q^2)^4}, \non \\&& \pi^{D/2} \int_0^\infty \frac{d\l}{\l^{(8-d)/2}} = \int \frac{d^D q}{(q^2)^2}, \quad \frac{\pi^D}{3} \int_0^\infty \frac{d\m d\theta d\r}{\Delta_2^{(8-d)/2} (\m,\r,\theta)}= \int \frac{d^D p d^D q}{(p^2)^2 (q^2)^2 (p+q)^2} \non \\ \eea
which relate the expressions involving Schwinger parameters to ones involving momentum integrals. These equalities simply follow by expressing the propagators in the momentum integrals in terms of Schwinger parameters and performing the momentum integrals. Thus this includes both one and two loop momentum integrals.

Now we evaluate the various momentum integrals in appropriate dimensions to obtain the simple poles in $\epsilon$. For the one loop integrals we have that
\bea \int \frac{d^D q}{(q^2)^2} = \frac{\pi^{D/2}}{\epsilon}, \quad 2\epsilon = 4-D, \non \\ \int \frac{d^D q}{(q^2)^3} = \frac{\pi^{D/2}}{2\epsilon}, \quad 2\epsilon = 6-D, \non \\ \int \frac{d^D q}{(q^2)^4} = \frac{\pi^{D/2}}{6\epsilon}, \quad 2\epsilon = 8-D.\eea   
For the two loop integral, we introduce Feynman parameters and perform the momentum integral. This gives us that
\be \label{T}\int \frac{d^D p}{(2\pi)^D} \frac{d^D q}{(2\pi)^D}\frac{1}{(p^2)^2 (q^2)^2 (p+q)^2} = \frac{\Gamma(5-D)}{(4\pi)^D} \int_0^1 dx dy dz  \frac{\delta(1-x-y-z)xy(1-y)^{5-D}}{[xy + z(1-z)]^{5-D/2}}.\ee
Once again we only need to isolate the pole term to obtain the logarithmic divergence. Setting $2\epsilon = 5-D$ and keeping only the pole term, we have that
\bea \int \frac{d^D p}{(2\pi)^D} \frac{d^D q}{(2\pi)^D}\frac{1}{(p^2)^2 (q^2)^2 (p+q)^2} &=& \frac{1}{2\epsilon (4\pi)^5} \int_0^1 dx dy dz  \frac{\delta(1-x-y-z)xy}{[xy + z(1-z)]^{5/2}} \non \\ &=& \frac{\pi}{3\epsilon (4\pi)^5}.\eea 

Thus adding the various contributions from \C{log1}, \C{log2}, \C{log3}, \C{log4} and \C{log5}, the total logarithmic divergence with moduli dependent coefficient functions, of the four graviton amplitude for the $D^4\mathcal{R}^4$ and $D^6\mathcal{R}^4$ amplitudes is given by
\bea \label{log}&&\mathcal{A}_{log} = \frac{\pi^D\kappa_{11}^2\kappa_D^4}{2\epsilon(2\pi)^{2D}}\s_2\mathcal{R}^4   \delta_{D,6}  \sum_{m_I}\int_0^\infty \frac{d\s}{ \s^{(4-d)/2}} e^{-\s {{\bf{m}}}^2} \non \\ &&+\frac{\pi^D\kappa_{11}^2\kappa_D^4}{24\epsilon(2\pi)^{2D}}\s_3\mathcal{R}^4  \delta_{D,8}\sum_{m_I}\int_0^\infty \frac{d\s}{ \s^{(4-d)/2}} e^{-\s {{\bf{m}}}^2} +  \frac{5\pi^D\kappa_{11}^2\kappa_D^4}{3\epsilon(2\pi)^{2D}}\delta_{D,4}\s_3\mathcal{R}^4 \non \\ &&\times \Big[\frac{1}{240}\sum_{m_I}'\int_0^\infty \frac{d\s}{ \s^{-d/2}} e^{-\s {{\bf{m}}}^2}  +\frac{\pi^{D/2}\kappa_D^2}{(2\pi)^D} \sum_{m_I,n_I}\int_0^\infty \frac{d\s d\l d\r }{\Delta_2^{(6-d)/2}(\s,\l,\r)}e^{-(\s {{\bf{m}}}^2 + \l {{\bf{n}}}^2 +\r{{\bf{(m+n)}}}^2)} \Big] \non \\ &&+ \frac{5\pi^{3D/2+1}\kappa_{11}^2\kappa_D^6}{3\epsilon(2\pi)^{3D}}\s_3\mathcal{R}^4 \delta_{D,5}\sum_{m_I}\int_0^\infty \frac{d\s}{\s^{(4-d)/2}} e^{-\s{\bf{m}}^2}\non \\ &&-\frac{5\pi^{3D/2}\kappa_{11}^2\kappa_D^6}{12\epsilon^2(2\pi)^{3D}}\s_3\mathcal{R}^4 \delta_{D,4}\sum_{m_I}'\int_0^\infty \frac{d\s}{\s^{(2-d)/2}} e^{-\s{\bf{m}}^2}.\eea
Thus using the expressions \C{11}, \C{10} and \C{D22} we see that the $D^4\mathcal{R}^4$ amplitude has a logarithmic divergence in 6 dimensions, with the coefficient involving the $\mathcal{R}^4$ amplitude. The $D^6\mathcal{R}^4$ amplitude has logarithmic divergences in 8 and 5 dimensions, with the coefficient involving the $\mathcal{R}^4$ amplitude as well. It also has a logarithmic divergence in 4 dimensions, with the coefficient involving the $D^4\mathcal{R}^4$ amplitude. All these divergences have a simple pole in $\epsilon$. In addition to these, the $D^6\mathcal{R}^4$ amplitude also has another logarithmic divergence in 4 dimensions, with the coefficient involving the $D^2\mathcal{R}^4$ amplitude, which is a double pole in $\epsilon$. 

We would now like to express the four graviton amplitude in terms of $\mathcal{E}_{\mathcal{R}^4}^{sugra}, \mathcal{E}^{sugra}_{D^4\mathcal{R}^4}$ and $\mathcal{E}^{sugra}_{D^6\mathcal{R}^4}$ such that comparing with \C{log} yields the required logarithmic divergences. In order to do so, we use the expression for $\mathcal{A}_{\mathcal{R}^4}$ in \C{final1} which leads to the term in the effective action
\be \label{action}S = \frac{1}{l_{11}^{D-8}}\int d^D x \sqrt{-G^{(D)}} \mathcal{V}_{d+1}\Big[ 4\zeta(2) + \mathcal{V}_{d+1}^{-3/(d+1)} E_{3/2}^{SL(d+1,\mathbb{Z})}\Big]\mathcal{R}^4\ee
in M theory compactified on $T^{d+1}$, where $G^{(D)}_{\m\n}$ is the M theory metric. Note that this leads to the genus one contribution $4\zeta(2)$ in 10 dimensions which is indeed the correctly normalized expression for the genus one amplitude. Now we would like to express \C{action} in terms of string theory variables. This is simply given by\footnote{The metric $G^{(D)}_{\mu\nu} = g^{A/B}_{\mu\nu} \equiv g_{\mu\nu}$ in the type IIA/B theory in the string frame of the IIA/B theory.}
\be S= \frac{1}{l_s^{D-8}}\int d^D x \sqrt{-g} \Big( \frac{l_s}{l_{11}}\Big)^{D-8}\mathcal{V}_{d+1}\Big[ 4\zeta(2) + \mathcal{V}_{d+1}^{-3/(d+1)} E_{3/2}^{SL(d+1,\mathbb{Z})}\Big]\mathcal{R}^4\ee  
in the string frame which we convert to the Einstein frame using
\be \label{sE} g_{\mu\nu} = g_d^{4/(D-2)} \hat{g}_{\mu\nu},\ee
where $\hat{g}_{\mu\nu}$ is the Einstein frame metric. This leads to the term in the effective action given by
\be S = \frac{1}{l_s^{D-8}} \int d^D x \sqrt{-\hat{g}} \mathcal{E}^{sugra}_{\mathcal{R}^4}\hat{\mathcal{R}}^4 ,\ee
in the Einstein frame where
\be \mathcal{E}^{sugra}_{\mathcal{R}^4} = \Big(\frac{l_D}{l_{11}}\Big)^{D-8} \mathcal{V}_{d+1}\Big[ 4\zeta(2) + \mathcal{V}_{d+1}^{-3/(d+1)} E_{3/2}^{SL(d+1,\mathbb{Z})}\Big],\ee
where $l_D$ is the $D$ dimensional Planck length defined by
\be \label{planck}l_D = g_d^{2/(D-2)} l_s.\ee
Thus we get that\footnote{Note that expressions like this have quantities both in the string frame ($\mathcal{R}^4$) as well as in the Einstein frame ($\mathcal{E}^{sugra}_{\mathcal{R}^4}$). Though we can convert everything to the Einstein frame, this is not necessary as our sole intention is to match these with \C{log}.}
\be \mathcal{A}_{\mathcal{R}^4} = \frac{\pi^3\kappa_{11}^2 l_D^6}{8(2\pi)^3} \mathcal{E}_{\mathcal{R}^4}^{sugra}\mathcal{R}^4\ee
on using
\be 2 \kappa_D^2 = (2\pi)^{D-3} l_D^{D-2}.\ee
Thus defining the four graviton amplitude\footnote{Thus the total logarithmic divergence is given by $\mathcal{A}_{log}$ in \C{log}.}
\be \mathcal{A} = \mathcal{A}_{\mathcal{R}^4} +\mathcal{A}_{D^4\mathcal{R}^4} +\mathcal{A}_{D^6\mathcal{R}^4},\ee
we have that
\bea \label{exp}\mathcal{A} = \frac{\pi^3\kappa_{11}^2 l_D^6}{8(2\pi)^3} \Big[\mathcal{E}^{sugra}_{\mathcal{R}^4} + \Big(\frac{l_D^2}{4}\Big)^2\s_2 \mathcal{E}^{sugra}_{D^4\mathcal{R}^4} + \Big(\frac{l_D^2}{4}\Big)^3\s_3 \mathcal{E}^{sugra}_{D^6\mathcal{R}^4}\Big]\mathcal{R}^4 \non \\ =  \frac{\pi^3\kappa_{11}^2 l_D^6}{8(2\pi)^3} \Big[\mathcal{E}^{sugra}_{\mathcal{R}^4} + \Big(\frac{l_s^2}{4}\Big)^2\hat\s_2 \mathcal{E}^{sugra}_{D^4\mathcal{R}^4} + \Big(\frac{l_s^2}{4}\Big)^3\hat\s_3 \mathcal{E}^{sugra}_{D^6\mathcal{R}^4}\Big]\mathcal{R}^4\eea
on pulling out a common factor from the entire amplitude. Note for example, that $S=-g^{\mu\nu}(k_1 + k_2)_\m (k_1 + k_2)_\n$ is the Mandelstam variable in the string frame, while $\hat{S} = -\hat{g}^{\mu\nu} (k_1 + k_2)_\mu (k_2 + k_2)_\nu$ is the Mandelstam variable in the Einstein frame. Thus
\be \hat{\s}_n = \hat{S}^n +\hat{T}^n +\hat{U}^n\ee  
which involves the Einstein frame metric.

As an elementary consistency check note that in 10 dimensions, the contributions from the genus zero amplitudes in $\mathcal{E}^{sugra}_{\mathcal{R}^4}$, $\mathcal{E}^{sugra}_{D^4\mathcal{R}^4}$ and $\mathcal{E}^{sugra}_{D^6\mathcal{R}^4}$ give us
\bea &&\mathcal{E}^{sugra}_{\mathcal{R}^4} + \Big(\frac{l_D^2}{4}\Big)^2\s_2 \mathcal{E}^{sugra}_{D^4\mathcal{R}^4} + \Big(\frac{l_D^2}{4}\Big)^3\s_3 \mathcal{E}^{sugra}_{D^6\mathcal{R}^4} \non \\ &&= 2\zeta(3) g^{-3/2} + \Big(\frac{ g^{1/2} l_s^2}{4}\Big)^2\s_2 \zeta(5) g^{-5/2}+ \frac{2}{3}\Big(\frac{g^{1/2} l_s^2}{4}\Big)^3\s_3 \zeta(3)^2 g^{-3} \non \\ &&= \Big[ 2\zeta(3) + \zeta(5)  \Big(\frac{\a'}{4}\Big)^2\s_2 +\frac{2}{3} \zeta(3)^2\Big(\frac{\a'}{4}\Big)^3\s_3\Big]g^{-3/2},\eea
which is precisely what is given by superstring perturbation theory (the overall remaining factor of $g^{-1/2}$ appears as an overall coefficient from the other factors in defining the string amplitudes). 

Thus from \C{11} and \C{exp} we have that
\be \mathcal{E}^{sugra}_{\mathcal{R}^4} = 2\pi^{D/2-3}l_D^{D-8}\sum_{m_I}\int_0^\infty \frac{d\s}{ \s^{(4-d)/2}} e^{-\s {{\bf{m}}}^2}, \ee
and from \C{D22} and \C{exp} we have that
\bea \mathcal{E}^{sugra}_{D^4\mathcal{R}^4} &=&\frac{32}{3} \pi^{D/2-3} l_D^{D-12}\Big[\frac{1}{240}\sum_{m_I}'\int_0^\infty \frac{d\s}{ \s^{-d/2}} e^{-\s {{\bf{m}}}^2}  \non \\ &&+\frac{\pi^{D/2}\kappa_D^2}{(2\pi)^D} \sum_{m_I,n_I}\int_0^\infty \frac{d\s d\l d\r }{\Delta_2^{(6-d)/2}(\s,\l,\r)}e^{-(\s {{\bf{m}}}^2 + \l {{\bf{n}}}^2 +\r{{\bf{(m+n)}}}^2)} \Big].\eea
Now following \C{exp} we define 
\be \mathcal{A}_{D^2\mathcal{R}^4} = \frac{\pi^3\kappa_{11}^2 l_D^{10}}{2^5 (2\pi)^3} \mathcal{E}^{sugra}_{D^2\mathcal{R}^4}\s_1 \mathcal{R}^4,\ee 
which using \C{10} gives us
\be \mathcal{E}^{sugra}_{D^2\mathcal{R}^4} = \frac{4}{15}\pi^{D/2-3}l_D^{D-10}\sum_{m_I}'\int_0^\infty \frac{d\s}{ \s^{(2-d)/2}} e^{-\s {{\bf{m}}}^2}.\ee

Hence we can express the divergent terms in $\mathcal{A}_{log}$ in \C{log} in terms of $\mathcal{E}^{sugra}_{\mathcal{R}^4}$, $\mathcal{E}^{sugra}_{D^2\mathcal{R}^4}$ and $\mathcal{E}^{sugra}_{D^4\mathcal{R}^4}$ giving us
\bea \label{Log}\mathcal{A}_{log} &=& \frac{\kappa_{11}^2 l_D^{10}\delta_{D,6}}{4^5\epsilon} \mathcal{E}^{sugra}_{\mathcal{R}^4} \s_2 \mathcal{R}^4 + \frac{\kappa_{11}^2 l_D^{12}}{4^6}\Big[\frac{\pi\delta_{D,8}}{3\epsilon} \mathcal{E}^{sugra}_{\mathcal{R}^4} +\frac{5\delta_{D,5}}{6\epsilon}\mathcal{E}^{sugra}_{\mathcal{R}^4}\non \\ && +\frac{5\delta_{D,4}}{2 \pi \epsilon} \mathcal{E}^{sugra}_{D^4\mathcal{R}^4} - \frac{25\delta_{D,4}}{16\pi^2\epsilon^2}\mathcal{E}^{sugra}_{D^2\mathcal{R}^4}\Big] \s_3\mathcal{R}^4. \eea

Finally comparing between \C{exp} and \C{Log}, we get the required logarithmic divergences with moduli dependent coefficients given by
\bea  \label{pole}\mathcal{E}_{D^4\mathcal{R}^4}^{non-an} &=&  \frac{\delta_{D,6}}{\epsilon}\mathcal{E}_{\mathcal{R}^4}, \non \\  \mathcal{E}_{D^6\mathcal{R}^4}^{non-an} &=& \frac{\pi\delta_{D,8}}{3\epsilon} \mathcal{E}_{\mathcal{R}^4} + \frac{5\delta_{D,5}}{6\epsilon} \mathcal{E}_{\mathcal{R}^4}+ \frac{5\delta_{D,4}}{2\pi \epsilon} \mathcal{E}_{D^4\mathcal{R}^4}- \frac{25\delta_{D,4}}{16\pi^2\epsilon^2}\mathcal{E}_{D^2\mathcal{R}^4},\eea
where we have sent $\mathcal{E}^{sugra} \rightarrow \mathcal{E}$ which is the U-duality completion. Note that $\mathcal{E}^{sugra}$ involves only finite expressions obtained after regularization. Hence the various amplitudes are defined along with the counterterm vertices if there are divergences (see for example the details of the regularization in appendix A).  

To obtain the expressions involving ${\rm ln} g_d$ we use the expressions derived in appendix B. All the simple poles in $\epsilon$ in \C{pole} apart from the one involving $\delta_{D,5}\mathcal{E}_{\mathcal{R}^4}$ arise from one loop divergences where we use \C{1loop}, while the one involving  $\delta_{D,5}\mathcal{E}_{\mathcal{R}^4}$ arises from a two loop divergence where we use \C{2loop}. The $1/\epsilon^2$ term arises from a two loop divergence, however the integral factorizes into the product of one loop integrals given in \C{log5}, and hence we use \C{1loop} for each of these factors. This leads to
\bea \label{final}&&\mathcal{E}_{D^4\mathcal{R}^4}^{non-an} =  \mathcal{E}_{\mathcal{R}^4} {\rm ln} g_4 \delta_{D,6}, \non \\  &&\mathcal{E}_{D^6\mathcal{R}^4}^{non-an} = \frac{2\pi}{9} \mathcal{E}_{\mathcal{R}^4}{\rm ln}g_2\delta_{D,8} + \frac{20}{9} \mathcal{E}_{\mathcal{R}^4}{\rm ln} g_5 \delta_{D,5}+ \frac{5}{\pi} \mathcal{E}_{D^4\mathcal{R}^4} {\rm ln}g_6\delta_{D,4}- \frac{25\delta_{D,4}}{4\pi^2}\mathcal{E}_{D^2\mathcal{R}^4} ({\rm ln} g_6)^2 \delta_{D,4}.\non \\ \eea
We now compare \C{final} and \C{value}. Except for the last term in $\mathcal{E}^{non-an}_{D^6\mathcal{R}^4}$ in \C{final}, the other terms agree precisely with those given in \C{value}. This extra contribution involves the coefficient function of an amplitude that is vanishing on--shell. Hence perturbative string amplitude calculations where such infrared divergent terms arise from the boundary of moduli space will not detect such terms. On the other hand the U--duality invariant equation for the $D^6\mathcal{R}^4$ coupling must include the contribution coming from this term, or else unitarity will be violated. In fact, the presence of coefficient functions of on--shell vanishing amplitudes as source terms for Poisson equations satisfied by the U--duality invariant couplings have been considered in~\cite{Basu:2008cf,Green:2008bf,Basu:2013goa,Basu:2013oka} in the context of theories with maximal supersymmetry.

Thus we see that the various logarithmically divergent contributions to these BPS interactions with moduli dependent coefficient functions that arise in specific dimensions can be determined directly from the detailed structure of multiloop amplitudes in supergravity. Of course, these coefficient functions have to be determined as a separate exercise.

Though our analysis has focused on the case of BPS amplitudes for simplicity, it can be generalized to some extent for non--BPS amplitudes that arise at higher orders in the derivative expansion of the effective action as well. This is facilitated by the known structure of the four graviton amplitude upto four loops~\cite{Bern:2009kd}\footnote{Certain contributions to non--BPS amplitudes from three and four loops in supergravity have been considered in~\cite{Basu:2014uba,Basu:2015dsa}.}. Even though these amplitudes are expected to receive contributions from all loops in supergravity, the presence of some of these logarithmically divergent contributions can be detected from loops at low orders in the supergravity expansion. Apart from the supergravity amplitudes, the structure of perturbative genus one string amplitudes resulting from the boundary of moduli space also point to the existence of such contributions to the non--BPS interactions~\cite{Basu:2016fpd}\footnote{The contributions to these BPS amplitudes from the boundaries of moduli space have been considered in~\cite{Basu:2015dqa,Pioline:2015nfa,Florakis:2016boz}.}. It would be interesting to understand such non--analytic terms in the effective action with duality invariant coefficients in theories with lesser supersymmetry.


{\bf{Acknowledgements:}} 


I am thankful to the theory group, IACS Kolkata for warm hospitality during the final stages of this work.

\appendix

\section{The $\mathcal{R}^4$, $D^2\mathcal{R}^4$ and $D^4\mathcal{R}^4$ coefficient functions from supergravity amplitudes}

In \C{value} and \C{add}, the coefficient functions of the various logarithmic terms involve the coefficient functions of the $\mathcal{R}^4$, $D^2\mathcal{R}^4$ and $D^4\mathcal{R}^4$ interactions. Hence we determine the contributions to them from supergravity. For the $D^2\mathcal{R}^4$ amplitude, we consider the one loop expression which we take to be its defining expression.   
 
At one loop for the $\mathcal{R}^4$ term we have that
\bea \mathcal{A}_{\mathcal{R}^4}^{(1)}  &=& \frac{\pi^{D/2}\kappa_{11}^2\kappa_D^2}{2(2\pi)^D} \mathcal{R}^4 \int_{1/\Lambda^2}^\infty d\s \s^{(d-4)/2} \sum_{m_I} e^{-\s {{\bf{m}}}^2}\non \\ &=& \frac{\pi^{11/2}\kappa_{11}^4}{2(2\pi)^D} \mathcal{R}^4 \sum_{\hat{m}^I} \int_0^{\Lambda^2} d\hat\s \sqrt{\hat\s} e^{-\pi^2 l_{11}^2 G_{IJ} \hat{m}^I \hat{m}^J \hat\s}\eea
where we have used the relation
\be \sum_{m_I}  e^{-\s  {{\bf{m}}}^2} = \frac{\pi^{(d+1)/2}l_{11}^{d+1} \mathcal{V}_{d+1}}{\s^{(d+1)/2}} \sum_{\hat{m}^I}e^{-\pi^2 l_{11}^2 G_{IJ} \hat{m}^I \hat{m}^J \hat\s}\ee
obtained by Poisson resummation, and defined $\hat\s = \s^{-1}$. The ultraviolet divergence has been cutoff at $\hat\s = \Lambda^2$. Thus we have that
\be  \label{div1} \mathcal{A}_{\mathcal{R}^4}^{(1)} =  \frac{\pi^3\kappa_{11}^4}{4(2\pi)^Dl_{11}^3} \mathcal{R}^4  \Big[ \frac{4\pi^{5/2}}{3} (\Lambda l_{11})^3 + \mathcal{V}_{d+1}^{-3/(d+1)} E_{3/2}^{SL(d+1,\mathbb{Z})}\Big],\ee
where we have defined the Eisenstein (or Epstein) series for $SL(d+1,\mathbb{Z})$ as\footnote{This is commonly referred to as $E^{SL(d+1,\mathbb{Z})}_{[1,0^{d-1}];s}$, where the subscripts refer to the $d$ Dynkin labels~\cite{Harish-Chandra,Langlands}. See~\cite{Green:2010wi} for example, for a review.}
\be  E_s^{SL(d+1,\mathbb{Z})} = \sum'_{\hat{l}^I} (\hat{G}_{IJ} \hat{l}^I \hat{l}^J)^{-s},\ee
where the sum excludes the contribution where all $\hat{l}^I =0$. Also note that
\be G_{IJ} = \mathcal{V}^{2/(d+1)}_{d+1} \hat{G}_{IJ},\ee
thus $G_{IJ}$ and $\hat{G}_{IJ}$ are the metrics on $T^{d+1}$ of volume $\mathcal{V}_{d+1}$ and $1$ respectively. Thus $\hat{G}_{IJ}$ depends on $d(d+3)/2$ shape moduli which parametrize the coset space $SO(d+1)\backslash SL(d+1,\mathbb{R})$.  

A one loop counterterm is added to \C{div1} which cancels the $\Lambda^3$ divergence, and the total finite contribution is given by~\cite{Green:1999pu}\footnote{In 8 dimensions, there is an additional ${\rm ln}\Lambda$ divergence which is removed by properly defining $E_{3/2}^{SL(3,\mathbb{Z})}$. This is also true for the $D^4\mathcal{R}^4$ amplitude in appropriate dimensions which we discuss later, and the logarithmic divergences are always taken to be renormalized.}
\be \label{final1}\mathcal{A}_{\mathcal{R}^4} =  \frac{\pi^3\kappa_{11}^4}{4(2\pi)^Dl_{11}^3} \mathcal{R}^4 \Big[ 4\zeta(2) + \mathcal{V}_{d+1}^{-3/(d+1)} E_{3/2}^{SL(d+1,\mathbb{Z})}\Big]\ee 
which we take to be the definition of the amplitude in quantum supergravity. This is one loop exact. 

The $D^2\mathcal{R}^4$ term is given by
\bea \label{add0}\mathcal{A}_{D^2\mathcal{R}^4} &=&  \frac{2\pi^{D/2}\kappa_{11}^2\kappa_D^2}{5!(2\pi)^D} \s_1 \mathcal{R}^4  \sum'_{m_I} \int_0^\infty d\s \s^{(d-2)/2} e^{-\s {{\bf{m}}^2}} \non \\ &=& \frac{2\pi^5\kappa_{11}^4}{5!(2\pi)^Dl_{11}\mathcal{V}_{d+1}^{1/(d+1)}}\s_1 \mathcal{R}^4   E_{1/2}^{SL(d+1,\mathbb{Z})}.\eea

At one loop for the $D^4\mathcal{R}^4$ term we have that
\bea \label{add1}\mathcal{A}_{D^4\mathcal{R}^4}^{(1)} &=&  \frac{\pi^{D/2}\kappa_{11}^2\kappa_D^2}{2\cdot 6!(2\pi)^D} \s_2 \mathcal{R}^4  \sum'_{m_I} \int_0^\infty d\s \s^{d/2} e^{-\s {{\bf{m}}^2}} \non \\ &=& \frac{\pi^{11/2}\kappa_{11}^4}{2\cdot 6!(2\pi)^D} \s_2\mathcal{R}^4 \sum'_{\hat{m}^I} \int_0^\infty d\hat\s \hat\s^{-3/2} e^{- \pi^2 l_{11}^2 G_{IJ} \hat{m}^I \hat{m}^J \hat\s} \non \\ &=& -\frac{\pi^7\kappa_{11}^4}{6!(2\pi)^D}\s_2 \mathcal{R}^4  l_{11} \mathcal{V}_{d+1}^{1/(d+1)} E_{-1/2}^{SL(d+1,\mathbb{Z})}.\eea
Note that in both \C{add0} and \C{add1} the sum in the first line excludes the contribution with all $m_I =0$, while the sum in the second line excludes the contribution with all $\hat{m}^I=0$. The $m_I =0$ term diverges as $\s \rightarrow \infty$ which is an IR divergence. The $\hat{m}^I=0$ diverges as $\hat\s \rightarrow 0$ which is again an IR divergence. They add up with other such terms to lead to non--local terms in the effective action. Hence the equality between the first two lines is this sense.   

We now consider the $D^4\mathcal{R}^4$ amplitude at two loops. 
We have that
\bea \label{eval}&&\mathcal{A}_{D^4\mathcal{R}^4}^{(2)} = \frac{\pi^D\kappa_{11}^2\kappa_D^4}{6(2\pi)^{2D}}\s_2\mathcal{R}^4  \sum_{m_I, n_I} \int_0^\infty \frac{d\s d\l d\r}{\Delta_2^{(6-d)/2}(\s,\l,\r)}  e^{-(\s {{\bf{m}}}^2 + \l {{\bf{n}}}^2 +\r{{\bf{(m+n)}}}^2)}\non \\ &&= \frac{\pi^{11}\kappa_{11}^4}{6(2\pi)^D} \s_2\mathcal{R}^4\sum_{\hat{m}^I, \hat{n}^I}\int_0^\infty d\hat\s d\hat\l d\hat\r \hat\Delta_2^{1/2}({\hat\s,\hat\l,\hat\r})e^{-\pi^2 l_{11}^2 G_{IJ}\Big(\hat\s \hat{m}^I \hat{m}^J + \hat\l \hat{n}^I \hat{n}^J +\hat\r( \hat{m}+\hat{n})^I (\hat{m}+\hat{n})^J\Big)/l_{11}^2}, \non \\ \eea
where
\be \hat\r = \frac{\r}{\Delta_2(\s,\l,\r)}, \quad \hat\s = \frac{\s}{\Delta_2(\s,\l,\r)}, \quad \hat\l = \frac{\l}{\Delta_2(\s,\l,\r)}\ee
and
\be \hat\Delta_2 (\hat\s,\hat\l,\hat\r)= \hat\s \hat\l + \hat\l \hat\r + \hat\r \hat\s = \Delta_2^{-1}(\s,\l,\r).\ee
We have also Poisson resummed using the relation
\bea &&\sum_{m_I, n_I} e^{-(\s {{\bf{m}}}^2 + \l {{\bf{n}}}^2 +\r{{\bf{(m+n)}}}^2)}\non \\  &&= \frac{(\pi l_{11}^2 \mathcal{V}_{d+1}^{2/(d+1)})^{d+1}}{\Delta_2^{(d+1)/2}(\s,\l,\r)}\sum_{\hat{m}^I, \hat{n}^I}e^{-\pi^2 l_{11}^2 G_{IJ}\Big(\hat\s \hat{m}^I \hat{m}^J + \hat\l \hat{n}^I \hat{n}^J +\hat\r( \hat{m}+\hat{n})^I (\hat{m}+\hat{n})^J\Big)/l_{11}^2}.\non \\ \eea

To evaluate \C{eval}, we further define~\cite{Green:1999pu}
\be \tau_1 = \frac{\hat\rho}{\hat\rho +\hat\lambda}, \quad \tau_2 = \frac{\sqrt{\hat\Delta_2(\hat\s,\hat\l,\hat\r)}}{\hat\rho+\hat\lambda}, \quad V_2 = l_{11}^2 \sqrt{\hat\Delta_2({\hat\s,\hat\l,\hat\r})},\ee
to get that
\be \label{SL2}\mathcal{A}_{D^4\mathcal{R}^4}^{(2)} =  \frac{\pi^{11}\kappa_{11}^4}{(2\pi)^D l_{11}^8} \s_2\mathcal{R}^4\sum_{\hat{m}^I, \hat{n}^I}\int_0^\infty dV_2 V_2^3 \int_{\mathcal{F}_2} \frac{d^2 \tau}{\tau_2^2} e^{-\pi^2 G_{IJ} (\hat{m}+\hat{n}\tau)^I (\hat{m}+\hat{n}\bar\tau)^J V_2/\tau_2},\ee
where $d^2 \tau= d\tau_1 d\tau_2$ and $\mathcal{F}_2$ is the fundamental domain of $SL(2,\mathbb{Z})$ defined by
\be \mathcal{F}_2 = \{ -\frac{1}{2} \leq \tau_1 \leq \frac{1}{2}, \tau_2 \geq 0, \vert \tau \vert^2 \geq 1\}.\ee 
Thus $V_2$ and $\tau$ parametrize the volume and complex structure of an auxiliary $T^2$. This integral which receives contributions from the various orbits of $SL(2,\mathbb{Z})$~\cite{Dixon:1990pc} have ultraviolet divergences that arise from the boundaries of moduli space. This gives us
\be \label{eval2} \mathcal{A}_{D^4\mathcal{R}^4}^{(2)} = \frac{\kappa_{11}^4}{(2\pi)^D} \s_2\mathcal{R}^4 \Big[ a\Lambda^8 + \frac{\pi^{13/2}\Lambda^3}{4l_{11}^5} \mathcal{V}_{d+1}^{-5/(d+1)} E_{5/2}^{SL(d+1,\mathbb{Z})}+ \frac{2\pi^4}{l_{11}^8} \mathcal{V}_{d+1}^{-8/(d+1)}\hat{E}_2^{SL(d+1,\mathbb{Z})}\Big] ,\ee 
where $a$ is an undetermined constant. The first term in \C{eval2} arising from $V_2 \rightarrow \infty$ receives a contribution from the zero orbit of $SL(2,\mathbb{Z})$), which yields the primitive two loop divergence of the 11 dimensional theory. This is regularized by a two loop counterterm leaving no finite remainder. 

The second term in \C{eval2} comes from $\tau_2 \rightarrow \infty$ which receives contribution from the degenerate orbit of $SL(2,\mathbb{Z})$. To evaluate this, we consider
\be \mathcal{J} = \frac{\pi^{11}}{2l_{11}^8}\sum_{\hat{m}^I, \hat{n}^I}' \int_0^\infty dV_2 V_2^3 \int_{\mathcal{F}_2} \frac{d^2 \tau}{\tau_2^2} e^{-\pi^2 G_{IJ} (\hat{m}+\hat{n}\tau)^I (\hat{m}+\hat{n}\bar\tau)^J V_2/\tau_2}\ee 
where not all $\hat{m}^I, \hat{n}^I$ can be zero. To isolate the contribution from $\tau_2 \rightarrow \infty$, we use the relation
\bea &&\Big[ \Delta_{SL(d+1,\mathbb{Z})} - \frac{d-1}{d+1} \Big(\mu^2 \frac{\p^2}{\p\mu^2} + (d+2) \m\frac{\p}{\p\m}\Big) \Big] e^{-\pi^2 G_{IJ} (\hat{m}+\hat{n}\tau)^I (\hat{m}+\hat{n}\bar\tau)^J V_2/\tau_2} \non \\ &&= \Delta_{\tau} e^{-\pi^2 G_{IJ} (\hat{m}+\hat{n}\tau)^I (\hat{m}+\hat{n}\bar\tau)^J V_2/\tau_2},\eea
where we have defined
\be \m = \mathcal{V}_{d+1}^{2/(d+1)}.\ee
Also the $SL(2,\mathbb{Z})$ invariant Laplacian is given by
\be \Delta_{\tau} = 4\tau_2^2\frac{\p^2}{\p\tau \bar\p\tau},\ee
while the $SL(d,\mathbb{Z})$ invariant Laplacian is given by~\cite{Obers:1999um}
\be \Delta_{SL(d,\mathbb{Z})} = \frac{1}{2} G_{IK} G_{JL} \frac{\p}{\p\tilde{G}_{IJ}}\frac{\p}{\p\tilde{G}_{KL}} +\frac{d+1}{2} G_{IJ} \frac{\p}{\p\tilde{G}_{IJ}} - \frac{1}{2d} \Big(G_{IJ} \frac{\p}{\p\tilde{G}_{IJ}}\Big)^2\ee
where
\be \tilde{G}_{IJ} = (1-\delta_{IJ}/2) G_{IJ}.\ee
We make use of the relation
\be \frac{\p G_{IJ}}{\p\tilde{G}_{KL}} = \delta^K_I \delta^L_J+\delta^K_J \delta^L_I.\ee
This leads to the differential equation
\be \Big[ \Delta_{SL(d+1,\mathbb{Z})} - \frac{d-1}{d+1} \Big(\mu^2 \frac{\p^2}{\p\mu^2} + (d+2) \m\frac{\p}{\p\m}\Big) \Big] \mathcal{J} = \frac{\pi^6\Lambda^3}{4l_{11}^5} \Gamma(7/2)\mu^{-5/2} E_{5/2}^{SL(d+1,\mathbb{Z})},\ee
which has the solution
\be \mathcal{J} = \frac{\pi^{13/2}\Lambda^3}{8l_{11}^5} \mathcal{V}_{d+1}^{-5/(d+1)} E_{5/2}^{SL(d+1,\mathbb{Z})}\ee
on using
\be \Delta_{SL(d,\mathbb{Z})} E_s^{SL(d,\mathbb{Z})} = \frac{s(d-1)(2s-d)}{d}E_s^{SL(d,\mathbb{Z})}.\ee
Thus this contribution yields a subdivergence which is cancelled by the one loop counterterm leaving a finite remainder.

Finally the last term in \C{eval2} is a finite contribution arising from the non--degenerate orbit of $SL(2,\mathbb{Z})$. We obtain the result by first performing the $\tau_1$ integral, and then performing the remaining ones by defining 
\be x= V_2/\tau_2, \quad y= V_2 \tau_2\ee
to perform the elementary integrals.
For this contribution we have defined another Eisenstein series for $SL(d+1,\mathbb{Z})$ as\footnote{This is commonly referred to as $E^{SL(d+1,\mathbb{Z})}_{[0,1,0^{d-2}];s}$.}
\be  \hat{E}_s^{SL(d+1,\mathbb{Z})} = \sum'_{d^{IJ}} (\hat{G}_{IK} \hat{G}_{JL} d^{IJ} d^{KL})^{-s},\ee
where $d^{IJ} = \hat{m}^I \hat{n}^J - \hat{m}^J\hat{n}^I$, and the sum excludes all $d^{IJ}=0$ .

Including the one loop counterterm contribution we get the finite expression
\be \label{add2}\mathcal{A}_{D^4\mathcal{R}^4}^{(2)} = \frac{\kappa_{11}^4}{(2\pi)^Dl_{11}^8} \s_2\mathcal{R}^4 \Big[\frac{\pi^{6}}{8} \mathcal{V}_{d+1}^{-5/(d+1)} E_{5/2}^{SL(d+1,\mathbb{Z})}+ 2\pi^4 \mathcal{V}_{d+1}^{-8/(d+1)}\hat{E}_2^{SL(d+1,\mathbb{Z})}\Big].\ee

Thus adding \C{add1} and \C{add2}, we get the complete expression
\bea \mathcal{A}_{D^4\mathcal{R}^4} &=& \frac{\kappa_{11}^4}{(2\pi)^Dl_{11}^8} \s_2\mathcal{R}^4\Big[ -\frac{\pi^7 l_{11}^9}{6!}   \mathcal{V}_{d+1}^{1/(d+1)} E_{-1/2}^{SL(d+1,\mathbb{Z})} + \frac{\pi^{6}}{8} \mathcal{V}_{d+1}^{-5/(d+1)} E_{5/2}^{SL(d+1,\mathbb{Z})}\non \\ &&+ 2\pi^4 \mathcal{V}_{d+1}^{-8/(d+1)}\hat{E}_2^{SL(d+1,\mathbb{Z})}\Big] \eea
which is the result in quantum supergravity.

\section{From $\epsilon$ poles to ${\rm ln} g_d$ at various loops}

We need to extract the normalization needed to go from the $\epsilon$ poles in dimensional regularization to the logarithm involving the string coupling at various loops in our analysis, which we now obtain. We consider specific types of loop amplitudes having UV divergences of the type we require to obtain the relation.

To start with consider the one loop amplitude 
\be L_1 = \int \frac{d^Dp}{(2\pi)^D} \frac{1}{(p^2)^n}\ee
which yields the pole term
\be L_1 = \frac{1}{(4\pi)^{D/2}\Gamma(n)\epsilon}\ee
in dimensional regularization where
\be 2n -D = 2\epsilon.\ee 
Alternatively doing the momentum integral using a momentum cutoff $\Lambda$, we get the logarithmic divergence
\be L_1 = \frac{{\rm ln} \Lambda^2}{(4\pi)^{D/2} \Gamma(n)}.\ee
Its contribution to the U-duality invariant coefficient function in the string effective action is determined by
\be {\rm ln}\Lambda^2 \rightarrow -{\rm ln}\Big(\frac{-S}{\Lambda^2}\Big) \rightarrow -{\rm ln}(-\alpha' S) \rightarrow -{\rm ln} (-\alpha' g_d^{4/(D-2)} \hat{S}) \rightarrow \frac{4}{D-2}{\rm ln} g_d,\ee
on using \C{sE}. In the first step, we have the UV divergent expression in supergravity in the string frame metric (recall that the M theory metric for the non--compact dimensions is the same as the string frame metric) where $S$ is a generic Mandelstam variable. This then determines the structure of the IR divergent term in the string calculation, which then converted to the Einstein frame gives the required logarithmic term. Thus the answer is obtained by a scaling argument and does not involve the details of the explicit momentum dependence.      

Thus at one loop we make the replacement
\be \label{1loop}\frac{1}{\epsilon} \rightarrow \frac{4}{D-2} {\rm ln}g_d.\ee

Next we consider the two loop integral
\bea L_2 &=& \int \frac{d^Dp}{(2\pi)^D} \int \frac{d^Dq}{(2\pi)^D}\frac{1}{(p^2)^{m_1} (q^2)^{m_2} ((p+q)^2)^{m_3}} \non \\ &=& \frac{\Gamma(\sum_i m_i)}{\prod_i \Gamma(m_i)}\int \frac{d^Dp}{(2\pi)^D} \int \frac{d^Dq}{(2\pi)^D}\int_0^1 dx dy dz\frac{\delta(1-x-y-z)x^{m_1 -1}y^{m_2 -1}z^{m_3-1}}{[xp^2 + y q^2 + z(p+q)^2]^{\sum_i m_i}}\non \\ \eea
where we have introduced Feynman parameters. These are cases where appropriate choices of $m_i$ give a simple pole in dimensional regularization\footnote{For example, \C{T} has $(m_1,m_2,m_3)= (2,2,1)$. The two loop primitive divergence for the $D^4\mathcal{R}^4$ amplitude in ordinary supergravity~\cite{Bern:1998ug} has $(m_1,m_2,m_3) = (3,3,1)$. On the other hand, for $(m_1,m_2,m_3) = (2,1,1)$, we have that
\be \int \frac{d^D p d^D q}{(p^2)^2 q^2 (p+q)^2} = \frac{\pi^D}{3}\int_0^\infty d\m d\r d\theta \frac{(\m+\r+\theta)}{\Delta_2^{D/2} (\m+\r+\theta)},\ee
which has a divergence of the form $1/\epsilon^2$ rather than a simple pole, as analyzed in \C{D} when $2\epsilon= 4-D$.}. Calculating the $\epsilon$ pole in dimensional regularization when
\be \sum_i m_i -D = 2\epsilon,\ee 
we get that
\be L_2 = \frac{1}{2\epsilon(4\pi)^D\prod_i \Gamma(m_i)} \int_0^1 dx dy dz\frac{\delta(1-x-y-z)x^{m_1 -1}y^{m_2 -1}z^{m_3-1}}{[x y + z(1-z)]^{\sum_i m_i/2}}.\ee
Alternatively calculating $L_2$ using a momentum cutoff we get the logarithmic divergence
\be L_2 = \frac{{\rm ln \Lambda^2}}{(4\pi)^D\prod_i \Gamma(m_i)}\int_0^1 dx dy dz\frac{\delta(1-x-y-z)x^{m_1 -1}y^{m_2 -1}z^{m_3-1}}{[x y + z(1-z)]^{\sum_i m_i/2}}.\ee
This leads to the replacement
\be \label{2loop}\frac{1}{\epsilon} \rightarrow \frac{2\cdot 4}{D-2} {\rm ln} g_d\ee
at two loops.

It is not difficult to generalize the analysis to L loops in $D$ dimensions, where there are $2n$ powers of momenta in the denominator of the integrand, and hence we define
\be \frac{2n}{L} - D = 2\epsilon.\ee  
We can compare the expressions obtained using dimensional regularization and momentum cutoff. Both methods produce the same final expression apart from a factor of ${\rm ln}\Lambda^2$ in momentum cutoff, and a factor of $\Gamma(n-LD/2) = 1/L\epsilon$ in dimensional regularization. This leads to the replacement
\be \label{Lloop}\frac{1}{\epsilon} \rightarrow \frac{4L}{D-2}{\rm ln} g_d\ee 
at $L$ loops. 

The expressions \C{1loop} and \C{2loop} are crucial in order to obtain \C{final}. Let us also make some further consistency checks to obtain the moduli independent logarithms for $\mathcal{E}_{\mathcal{R}^4}^{non-an}, \mathcal{E}_{D^4\mathcal{R}^4}^{non-an}$ and $\mathcal{E}_{D^6\mathcal{R}^4}^{non-an}$ in \C{value} in 8, 7 and 6 dimensions respectively. 

For $\mathcal{E}_{\mathcal{R}^4}$, the one loop box diagram gives the pole
\be \mathcal{E}_{\mathcal{R}^4}^{non-an} = \frac{2\pi}{\epsilon},\ee 
where $2\epsilon = 8-D$, leading to
\be \label{one}\mathcal{E}_{\mathcal{R}^4}^{non-an} = \frac{4\pi}{3}{\rm ln}g_2 \delta_{D,8}. \ee
For $D^4\mathcal{R}^4$, the two loop primitive divergence has a simple pole~\cite{Bern:1998ug} given by
\be \mathcal{E}_{D^4\mathcal{R}^4}^{non-an} = \frac{2\pi^2}{3\epsilon}\ee
where $2\epsilon = 7-D$, leading to
\be  \label{two}\mathcal{E}_{D^4\mathcal{R}^4}^{non-an} = \frac{16\pi^2}{15}{\rm ln} g_3\delta_{D,7}.\ee 
Finally for $D^6\mathcal{R}^4$, the three loop primitive divergence has a pole~\cite{Bern:2008pv} leading to
\be \mathcal{E}_{D^6\mathcal{R}^4}^{non-an} = \frac{5\zeta(3)}{3\epsilon}\ee
where $2\epsilon = 6-D$, leading to
\be  \label{three}\mathcal{E}_{D^4\mathcal{R}^4}^{non-an} = 5\zeta(3){\rm ln} g_4\delta_{D,6}.\ee 
Now the expressions for the logarithmic contributions in \C{one}, \C{two} and \C{three} precisely match those in \C{value}. These terms that arise from primitive divergences in multiloop supergravity have also been considered in~\cite{Green:2010sp}. 


\providecommand{\href}[2]{#2}\begingroup\raggedright\endgroup

\end{document}